\documentclass[10pt]{article}

\usepackage[utf8]{inputenc}
\usepackage[L7x]{fontenc}
\usepackage{multirow}
\usepackage[american]{babel}
\usepackage{lmodern}
\usepackage{longtable}
\usepackage{enumerate}
\usepackage{hyperref}
\usepackage{graphicx}
\usepackage{epstopdf}
\usepackage{a4wide}
\usepackage[usenames]{color}
\usepackage{soul}
\usepackage[all]{xy}
\usepackage{array}
\usepackage{tabularx}
\usepackage[table]{xcolor}
\usepackage{booktabs}
\usepackage{colortbl}
\usepackage{verbatim}
\usepackage{rotating}
\usepackage{xtab}
\usepackage{lmodern}
\usepackage[top=2cm, bottom=2cm, left=3cm, right=3cm, footskip=1cm, a4paper]{geometry}
\usepackage{amssymb, amsmath, amsthm}
\usepackage{etoolbox}
\usepackage{morefloats}
\usepackage{footnote}
\usepackage[toc,page]{appendix}
\usepackage[bottom, splitrule]{footmisc}
\usepackage{float}
\usepackage[font=small]{caption}
\usepackage[round, sort, numbers, authoryear]{natbib} 
\usepackage{chngcntr}
\usepackage{dsfont}
\usepackage{tipa}
\usepackage{covington}
\usepackage[nottoc]{tocbibind}
\usepackage{enumitem}

\usepackage{array,colortbl,xcolor}
\def\boxit#1{%
	\smash{\color{black}\fboxrule=1.2pt\relax\fboxsep=1.45pt\relax%
		\llap{\rlap{\fbox{\vphantom{0}\makebox[#1]{}}}~}}\ignorespaces
}

\newenvironment{sciabstract}{%
\begin{quote} \small}
{\end{quote}}

\newcounter{lastnote}

\newcommand\numberthis{\addtocounter{equation}{1}\tag{\theequation}}

\addto{\captionsenglish}{}
\addto{\captionsbritish}{}
\addto{\captionsamerican}{}

\theoremstyle{plain}

\begin{document} 

\title{Sparse structures with LASSO through Principal Components: forecasting GDP components in the short-run}

\author
{
	Saulius Jokubaitis$^{1}$, Dmitrij Celov$^{1}$, Remigijus Leipus$^{1}$\\
	\\
	{\small \sl $^{1}$Institute of Applied Mathematics, Faculty of Mathematics and Informatics, Vilnius University, }\\
	{\small \sl Naugarduko 24, Vilnius LT-03225, Lithuania}\\
}

\date{}

\maketitle 

\begin{sciabstract}
This paper aims to examine the use of sparse methods to forecast the real, in the chain-linked volume sense, expenditure components of the US and EU GDP in the short-run sooner than the national institutions of statistics officially release the data. We estimate current quarter nowcasts along with 1- and 2-quarter forecasts by bridging quarterly data with available monthly information announced with a much smaller delay. We solve the high-dimensionality problem of the monthly dataset by assuming sparse structures of leading indicators, capable of adequately explaining the dynamics of analyzed data. For variable selection and estimation of the forecasts, we use the sparse methods – LASSO together with its recent modifications. We propose an adjustment that combines LASSO cases with principal components analysis that deemed to improve the forecasting performance. We evaluate forecasting performance conducting pseudo-real-time experiments for gross fixed capital formation, private consumption, imports and exports over the sample of 2005-2019, compared with benchmark ARMA and factor models. The main results suggest that sparse methods can outperform the benchmarks and to identify reasonable subsets of explanatory variables. The proposed LASSO-PC modification show further improvement in forecast accuracy.
  
  
  
\end{sciabstract}

\section*{Introduction}
Information on the current state of the economy is crucial for various economic agents and policymakers since the choice of the appropriate policy measures relies on the current knowledge of the macroeconomic situation in the country. Although there are some higher frequency indicators, covering many aspects of the economy, quarterly national accounts still play a pivotal role in guiding economic policy decisions. Unfortunately, the official release of GDP and its components occurs with a considerable delay after the reference period. For instance, the flash estimates of the US and EU GDP are released one month after the reference period, where only the supply side, i.e. production of goods and services, is covered. While the demand side reflected in GDP accounted by expenditure approach is released with an even longer two months delay. Bearing in mind economic policy implementations lags, the apt nowcasting and short-run forecasting tools are crucial for the regular monitoring and efficient countercyclical policy implementation. The latter is vitally important during severe downturn times like the global financial crisis of 2007-2008 followed by the sovereign debt crisis in the EU, and the current COVID-19 led global lockdown.

At the same time, the agencies of national statistics announce various short-term monthly indicators, such as business or consumer surveys, the industrial production, retail or external trade indicators with a much shorter delay. The monthly data allow getting a preliminary picture of how the current economic activity evolves in various sectors of the economy. One way of using higher frequency information is by conducting the fundamental analysis. However, the quarterly and annual national accounts data remain the core in guiding most of the economic policy decisions, e.g. in the area of fiscal rules. The need to extract the underlying signals from the available higher frequency data and obtain the lower frequency national accounts data sooner than the official release by the agencies of national statistics led to the development of several econometric tools examined in this paper. 

The literature suggests many different methods for bridging the higher and lower frequency data, the mainstream of which are the factor models and their modifications (\cite{BANBURA2013195}) altered to use all available information. Therefore, such methods possess a density feature.
On the other hand, the sparsity feature of the data hints that only a small subset of all the available information might be sufficient for adequate and timely estimation of the GDP or its components (\cite{Bai2008304}). 

In this paper, we further explore the sparse structure approach under a large amount of monthly economic indicators, assuming that only a small subset of them is significant and valuable in forecasting macroeconomic variables. Therefore, the underlying problem is the optimal selection of useful predictors. \cite{Bai2008304} find that large amounts of high-frequency data could often considerably worsen predictive performance, raising a question of how much of high-frequency data required for good predictions? On top of that, \cite{RollingWindow} among others show that assuming a sparse structure of the underlying data improve the quality of short-run forecasts when the final set of available information is refined by supervised selection before the application of bridge equations modelling approach. 

Recently a rapid growth in popularity among the practitioners and the academics is seen applying the \textit{Least Absolute Shrinkage Selection Operator} (LASSO) method pioneered by \cite{Tibshirani}. This approach employs the supervised variable selection for modelling showing great potential when used for both variable selection and the generation of accurate forecasts of economic data in the short run. For this reason, we, first, review the salient features of LASSO and its recent modifications: the Square-Root LASSO, the Adaptive LASSO and the Relaxed LASSO. Second, we propose a combination of the LASSO modifications with the method of Principal Components seeking to extract the significant underlying information with greater accuracy. We evaluate the empirical performance of the models by conducting pseudo-real-time short-run forecasting experiments of real, in chain-linked volumes sense, US, France, Germany, Spain and Italy GDP components accounted by expenditure approach: the Gross Fixed Capital Formation (GFCF), Private Final Consumption Expenditure (PFCE), Imports and Exports of goods and services. During the validation exercise, we estimate forecasts at four different forecasting horizons: the backcast of a previous quarter, the nowcast of the coinciding quarter and 1- and 2-quarter forecasts over the sample of 2005-2019. We find evidence of further improvement of the forecasting performance while applying proposed LASSO-PC method.

The motivation for looking at the demand breakdown but not the GDP itself is twofold. First, there is evidence in the literature (\cite{Dreschel}), that forecasting GDP by the bottom-up approach can lead to more accurate forecasts than direct forecasts. This claim follows from the fact that we can examine the different underlying and sparse structures of the subcomponents by modelling them separately. For example, investments and international trade are much more volatile than aggregate GDP, while private consumption is typically smoother than total activity (\cite{OBES:OBES092}). Therefore, in this paper, we aim to study how do different models compare in forecasting variables that are behaving so differently over the business cycle. Second, it is evident that the business cycle behaviour of the aggregate GDP is very different from that of its subcomponents. For example, investments tend to trough before GDP, while consumption only takes momentum when an expansion is well underway, peaking only after the cycle (\cite{CASTRO2010347}). Therefore, forming models for the subcomponents of GDP can not only improve the accuracy of the final aggregate GDP but also act as a complement to the final forecasts. Recovering information on the main drivers of the economic activity by itself may allow for a more accurate read on the cyclical phase by the economic agents. Additionally, forming a different model for each of the subcomponents admits the inclusion of different sets of predictors providing a richer story behind the specified equation. 

The paper is organized as follows. Section \ref{lasso} presents a detailed overview of the LASSO and some of its popular modifications found in the literature.  Section \ref{PCA} describes the proposed modification of combining the LASSO modifications with principal components, and further, expands into Section \ref{sparsePCA} where the possible alterations of the approach are discussed. For an empirical study, Section \ref{duomenys} describes the design of the information set and the settings for all models used in the pseudo-real-time exercise, while Section \ref{chapter:PR} presents the results of the pseudo-real-time experiment, where the backcast, nowcast and 1- and 2-quarter forecasts are estimated. The forecasting performance of the models is then compared with the performance of the benchmark ARMA and factor models.

\section{Short overview of the LASSO methods}
\label{lasso}
Throughout the paper we assume that $Y:=(Y_1, \ldots, Y_n) $ is the modelled variable with $n$ available observations; $p$ is the total number of explanatory variables $X_j \in \mathbb{R}^{1 \times n}$ used in the search space, where $j = 1, \ldots, p$. We model $Y =  \beta X + \varepsilon$, where $\beta = (\beta_1, \ldots, \beta_p)$, $\varepsilon = (\varepsilon_1, \ldots, \varepsilon_n)$ is the error term, and $X = (X_1', \ldots, X_p')' \in \mathbb{R}^{p \times n}$ is the explanatory variable matrix. $||v||_q := ( \sum_{j = 1}^{d}|v_j|^q )^{1/q}$ denotes the $\ell_q$-norm, $q \geq 1$, and $||v||_0 := \sum_{j=1}^{d} \mathds{1}\{v_j \neq 0\}$ for $q = 0$, where $v \in \mathbb{R}^{d}$ for some $d \in \mathbb{N}_{+}$. 

\subsection*{LASSO}

The LASSO method (\cite{Tibshirani}) is one of the possible solutions dealing with $p \gg n$ problem, where the ordinary least squares (OLS) method is infeasible due to the `curse-of-dimensionality', when the number of estimated parameters is much larger than the sample size. LASSO attracts a lot of attention in the literature due to its favourable strictly convex optimization problem, the solution path of which can be effectively estimated by using the Least angle regression (LARS) algorithm (\cite{efron2004}).

LASSO is an $\ell_1$-norm penalized least squares algorithm that solves the following optimization problem: 
\begin{equation}
\label{LASSO}
\hat \beta_{\text{LASSO}} = \underset{\beta}{\arg\min} ~(Y - \beta X )(Y - \beta X )' + \lambda || \beta ||_1,
\end{equation}
where the hyperparameter $\lambda \in (0, \infty)$ is fixed. With the value of $\lambda$ tending to $\infty$, the estimated coefficients are shrunken towards zero, and with a sufficiently large value of $\lambda$ some of them are estimated as 0 due to the properties of the $\ell_1$-norm retaining the most significant features. Besides, there exists such $\lambda$, starting from which none of the variables are included in the model as significant. However, decreasing the value of $\lambda$ the significant variables are sequentially included to the modelled regression. That is, in relation to the hyperparameter $\lambda$, LASSO procedure can be interpreted as a stepwise regression with an additional shrinkage of the estimated model's parameters towards zero. Also, due to the shrinkage of the estimated coefficients it is often possible to increase the accuracy of the forecasts, since the shrunken coefficients are able to reduce the variance of the forecasts by more than increasing the bias implying the solution to bias-variance trade-off.

A lot of attention in the literature is given particularly to the variable selection aspect of the LASSO. However, in a typical real-world case, we cannot fully expect both consistent variable selection and parameter estimation. In the following text, we will refer to these properties as the Oracle Properties, defined in  \cite{FanLi}.  For more detailed asymptotic results the interested reader should refer to, e.g., \cite{knight2000}, \cite{Zhao} and \cite{doi:10.1198/016214506000000735}.

\subsection*{Adaptive LASSO}
\label{adaLasso}

In the literature, many modifications of the LASSO aim to overcome various shortcomings of the original method. One of the most popular is the Adaptive LASSO, allowing to define weights for each explanatory variable used in the model:

\begin{equation}
\hat\beta_{\text{adaLASSO}} = \underset{\beta}{\arg\min} (Y -  \beta X)(Y - \beta X)' + \lambda  w' |\beta|,
\label{adaLASSO}
\end{equation}
where $|\beta| := (|\beta_1|, \ldots, |\beta_p|)'$ and $w = (w_1, \ldots, w_p)'$  is a vector of fixed weights. \cite{doi:10.1198/016214506000000735} proves that when the weight vector $w$ is data-driven and appropriately chosen, the Adaptive LASSO is able to achieve the Oracle Properties. The literature often suggests to choose the weights as $\hat w_j := 1/| \hat \beta_j^{*}|^{\gamma}$, $j = 1, \ldots, p$,  $\gamma > 0 $, where  $\hat\beta^{*}$ is $\sqrt{n}$-consistent estimator of $\beta$. When $p \leq n$, a natural choice could be $\hat\beta := \hat\beta_{\text{OLS}}$. However, when the OLS estimates are infeasible ($p > n$) or the data is strongly multicollinear, it is suggested to replace the $\hat\beta_{\text{OLS}}$ with $\hat\beta_{\text{ridge}}$, where coefficients are estimated by the Ridge regression defined by the problem similar to LASSO (\ref{LASSO}), where the $\ell_1$-norm of the imposed penalty is replaced with the $\ell_2$-norm. It is argued that with the sample size increasing, the weights for the insignificant variables should tend to infinity, while the weights of the significant variables should converge to some finite non-zero constant. This way, the method allows for an asymptotically unbiased simultaneous estimation of large coefficient and small threshold estimates.

For more detailed discussion on the possible choice of the weight vector the reader should refer to \cite{10.2307/24308572}, \cite{RePEc:rio:texdis:636}, \cite{Doubly}, \cite{FOR:FOR2403}. Additionally, \cite{RePEc:rio:texdis:636} show the asymptotics of the Adaptive LASSO on high dimensional time series, where the Oracle Properties hold under certain restrictions for both non-Gaussian and conditionally heteroscedastic data, which are traits, often found in practice.

Finally, \cite{Doubly} observes that the Adaptive LASSO can be effectively performed by employing the LARS algorithm: let $\hat W := \text{diag}(\hat w_1, \ldots, \hat w_p)$, then the optimization problem of the Adaptive LASSO (\ref{adaLASSO}) can be rewritten  as:

\begin{align*}
\hat\beta_{\text{adaLASSO}} &= \underset{\beta}{\arg\min} (Y - \beta  \hat W \hat W^{-1} X)(Y - \beta  \hat W \hat W^{-1} X)' + \lambda ||\beta  \hat W||_1 \\ &= \underset{\tilde\beta}{\arg\min} (Y - \tilde\beta\tilde X  )(Y - \tilde\beta\tilde X  )' + \lambda ||\tilde \beta||_1, \numberthis \label{adaLASSO2}
\end{align*}
where $\tilde X =  \hat W^{-1} X$, and $\tilde \beta = \beta  \hat W $ (i.e., $\tilde \beta_j = \hat w_j \beta_j, \forall j$), therefore all of these parameters can be effectively estimated by using the LARS algorithm just as the ordinary LASSO method.

\subsection*{Relaxed LASSO}

Another popular modification of the LASSO, dealing with some of its shortcomings, is the Relaxed LASSO (\cite{Meinshausen2007374}). The main idea of this method is to separate the selection of the significant variables and the estimation of the model's coefficients by introducing an additional hyperparameter $\phi$. Let $\mathcal{M}_{\lambda} := \{ 1 \leq j \leq p ~ : ~ \hat \beta_{j}^{\lambda} \neq 0 \}$ denote the set of variables, preselected by the LASSO method under a certain fixed value of $\lambda$. Then the Relaxed LASSO is estimated as:

\begin{equation}
\label{reLASSO}
\hat\beta_{\text{reLASSO}} = \underset{\beta}{\arg\min ~} n^{-1}\big(Y - \{\beta \cdot \mathds{1}_{\mathcal{M}_{\lambda}}\}X\big)\big(Y - \{\beta \cdot \mathds{1}_{\mathcal{M}_{\lambda}}\}X\big)'  + \phi \lambda ||\beta||_1,
\end{equation}
where $\lambda \in [0, \infty)$ and $\phi \in (0, 1]$, with $\mathds{1}_{\mathcal{M}_{\lambda}}$ being the indicator function, returning the value of 1 for those two variables, that were selected by the LASSO as significant under a fixed $\lambda$. That is, for a fixed $\lambda$, the following holds for the set of significant variables $\mathcal{M}_{\lambda} \subset \{ 1, \ldots, p \}$:
\begin{equation}
\{ \beta \cdot \mathds{1}_{\mathcal{M}_{\lambda}}\}_{k} = \left\{ \begin{array}{lc}0&k\notin \mathcal{M}_\lambda,\\
\beta_k&k \in \mathcal{M}_{\lambda},\\\end{array} \right.
\label{bb}
\end{equation}
for every  $k \in \{1, \ldots, p\}$. In this way, the selection of significant variables is performed by using the ordinary LASSO and estimating only the hyperparameter $\lambda$, while the appropriate estimation of the model's parameters and the amount of shrinkage applied is refined by using a second hyperparameter $\phi$. When  $\phi = 1$, the estimator coincides with the case of the ordinary LASSO, that is, no correction of the estimated coefficients is performed, while when $\phi$ approaches zero the estimator converges to Post-LASSO estimator (\cite{belloni2013}).

\cite{Meinshausen2007374} prove that due to such separation of the variable selection, the consistent estimates of the model's coefficients are obtained with the usual $\sqrt{n}$ rate of convergence, independently from the growth rate of the available information set. 

\subsection*{Square-Root LASSO}

Another recent modification of the LASSO is the Square-Root LASSO (\cite{doi:10.1093biometasr043}). The authors propose modifying the original formulation of the LASSO problem (\ref{LASSO}) by taking the square root of the residual sum of squares term, as defined by the equation (\ref{sqLASSO}):
\begin{equation}
\label{sqLASSO}
\hat\beta_{\text{sqrtLASSO}} =\underset{\beta}{\arg\min}~n^{-1/2}\left((Y-\beta X)(Y-\beta X)'\right)^{1/2} + \lambda ||\beta||_1.
\end{equation}

The attractiveness of the method follows the original idea for LASSO presented in \cite{bickel2009}, where rate-optimal value of $\lambda = \sigma \cdot 2 \sqrt{2 \log(pn)/n} $, depends on unknown value of $\sigma$. \cite{doi:10.1093biometasr043} show that for the Square-Root LASSO the rate-optimal penalty level reduces to $\lambda = \sqrt{2 \log(pn)/n}$, which makes it having no user-specified parameters and therefore tuning free.

\subsection*{Fast Best Subset Selection}

Regularization methods like LASSO try to approximate the computationally infeasible best subset selection algorithm. Recently, \cite{hazimeh2018fast} proposed a Fast Best Subset Selection (FBSS) realization by introducing several proxy algorithms and various heuristics, making the combinatorial search feasible. The authors propose the following optimization problem:

\begin{equation}
\hat \beta_{L0Lq} = \underset{\beta}{\arg\min} \frac{1}{2}(Y - \beta X)(Y - \beta X)' + \lambda_0 || \beta ||_0 + \lambda_q || \beta ||_{q}^{q}
\end{equation}
where $q \in \{1,2\}$ determines the type of additional regularization. Here the $\lambda_0$ controls the number of variables selected, while the $\lambda_q$ controls the shrinkage imposed on the estimates. Setting $q=0$ in this paper reduces to best subset selection. However, the authors recommend additional shrinkage due to observed overfitting in certain cases.

\section{Principal Components and LASSO}
\label{PCA}
In this paper we propose a combination of the aforementioned LASSO modifications together with principal components in order to preserve specific strengths and to minimize the possible shortcomings for each of the methods combined.

We follow the arguments of \cite{Bai2008304}, who show that the use of targeted predictors helps achieve significantly better forecasts of macroeconomic data using factor models. Instead of the usual approach to factor model forecasting, where the principal components follow from the full data set, the authors suggest using only a subset selected by a chosen hard/soft thresholding algorithms. In this way, an unsupervised algorithm becomes supervised one, because the choice of the targeted predictors now depends on the predicted variable. Following these arguments, we propose employing the LASSO, Adaptive LASSO and Square-Root LASSO for variable selection. These modifications deemed to more accurately select variables due to their known asymptotic properties under $p \gg n$ allowing for many highly correlated variables. From here on, let us assume that $X^{\lambda} \in \mathbb{R}^{q \times n}$ is a preselected matrix of significant variables under a fixed $\lambda$ by some variant of LASSO, where $0 < q  \leq n $, and is scaled and centered. 

Following these arguments, we present the proposed algorithm in Section \ref{lassopca}, and follow with the explanation of the motivation behind each step in Section \ref{lassopcamotivation}. In Section \ref{sparsePCA} we discuss further fine-tuning of the proposed algorithm. 

\subsection{LASSO-PC}
\label{lassopca}
We propose the following six-step algorithm:

\begin{enumerate}[label=(\alph*)]
	\item Use LASSO or some of its variant\footnote{The interested reader may wish to try combining additional methods for the better choice of candidate variables, e.g., by employing the Rolling Window variable selection approach by \cite{RollingWindow}, or similar.} to select a subset of candidate indicators $X^{\lambda} \in \mathbb{R}^{q \times n}$ from the full dataset $X \in \mathbb{R}^{p \times n}$, under a certain fixed $\lambda$, were $0 < q \leq n$, and the data is assumed as scaled and centered.
	\item Rotate $X^{\lambda}$ by principal components to $ F = L ' X^{\lambda}$, where $L \in \mathbb{R}^{q \times q}$ is a rotation matrix and $F \in \mathbb{R}^{q \times n}$ is a principal components matrix.\footnote{Besides the classical PCA, one may consider alternatives, e.g., by fine-tuning the angles or scale of the rotation (briefly discussed in Section \ref{sparsePCA}), or using another viable alternative, like Independent Component Analysis (see, for example, \cite{KIM2018339}).} 
	\item Forecast each series in $X^{\lambda}$ by an appropriate ARIMA model. In our case, we forecast at the monthly frequency, and forecasts are aggregated back to quarterly frequency.
	\item Aggregate time series forecasts to their factor representation using the rotation matrix $L$, i.e., $\hat F_{n+h}^{*} = L' \hat X_{n+h}^{\lambda}$, for $h = 1, 2, \ldots.$
	\item Use LASSO to estimate the coefficients in $Y_n = \beta \hat F_{n} + \varepsilon_n$.\footnote{Further, one may want to consider expanding the estimation by estimating the coefficients in $Y_n = \beta (\hat F_{n}', X'^{\lambda}_n)' + \varepsilon_n$. The motivation behind this step is discussed in Section \ref{sparsePCA}.}
	\item Produce the forecasts $\hat Y_{n+h} = \hat \beta \hat F_{n+h}$. 
\end{enumerate}

\subsection{Motivation}
\label{lassopcamotivation}
Since we are interested in modelling macroeconomic data, it is likely, that significant correlation will be observed, with some of the variables possibly even being nested.
Therefore, our proposed modification stems from the initial variable selection step of the Relaxed LASSO. However, instead of a direct parameter re-estimation of the selected variables, as would be done on the second step of the Relaxed LASSO, we propose rotating the data using the principal components methodology. In other words, we propose extracting the main latent factors $ F = L' X^{\lambda}$, where $L \in \mathbb{R}^{q \times q}$ is a rotation matrix and $F \in \mathbb{R}^{q \times n}$ is a principal components matrix. 

The main idea here is to extract the underlying information from the data as orthogonal latent factors and to model them instead. Due to probable inter-correlation and the supervised preselection done in the first step -- it is likely that such data capture prevalent signals, driving the particular market or the economic sector in question. If we assume those signals being the main reason for macroeconomic growth, it is a good idea to include them directly instead of the original data. 

As for the parameter estimation, we expand on the idea, defined by equation (\ref{adaLASSO2}),  and estimate:
\begin{align*}
\hat\beta_{\text{LASSO-PC}} &= \underset{\beta}{\arg\min ~}  (Y - \beta L L' X^{\lambda} )(Y - \beta L L' X^{\lambda} )' + \lambda ||\beta L||_1 \\ &= \underset{\tilde\beta}{\arg\min ~}  (Y - \tilde \beta F)(Y - \tilde \beta F )' + \lambda ||\tilde \beta||_1, \numberthis \label{fLASSO}
\end{align*}
\noindent
since $LL' = I$ holds by the definition of principal components, $\tilde \beta = \beta L$; all of which can be efficiently estimated using the LARS algorithm. It can be noted that $LL' = I$ holds for any $\tilde q \leq q$, so it is feasible to remove the redundant components, which explain very little of the total variance of the data and have very small loading coefficients if there are any.

Our proposed approach differs from the one suggested by  \cite{Bai2008304}, first of all, because the number of significant factors is selected not by the usual selection, based on various information criteria (such as \textit{Akaike}, \textit{Schwarz}, t-statistics from OLS and similar), but by using the soft-thresholding LASSO approach. That is, both the selection of significant factors and the shrinkage of estimated parameters applied simultaneously deemed to improve the forecasting accuracy, which is the main objective of short-run forecasting. 

The main strength of such an approach is when dealing with a large amount of data,
driven by one or only a few leading factors. In that case, the principal component transformation would allow us to extract those latent factors and estimate only the predictive ones using the LASSO. The final coefficient vector ${\hat{\beta}^{*}}:=  {\hat{\tilde \beta}} L$ would be comprised of the same non-zero variables just as it would be in the classic (Relaxed or Adaptive) LASSO case, but the estimated parameters would be set according to the significance and predictive performance in the latent space, rather than the direct one. Noteworthy, such transformation would act as a filter, distinguishing the essential underlying signals from the data and possibly allowing for more accurate forecasting performance.

Second, in contrast to \cite{Bai2008304} and other similar factor forecasting related literature, we propose to base the final forecasts on the predictions of individual variables $X^{\lambda}_j$, $j=1, \ldots, q$, rather than on the predicted significant factors $F_j$. That is, if we denote $X^{\lambda}_t = (X^{\lambda}_{1,t}, \ldots, X^{\lambda}_{q,t})'$ as the data at a time moment $t \in \{1, \ldots, n\}$, then for every $h > 0$, the forecasts $\hat F_{n+h}^{*} = (\hat F_{1,n+h}, \ldots, \hat F_{q,n+h})' $ can be calculated as 
\[
\hat F_{n+h}^{*} = L' \hat X_{n+h}^{\lambda}   = L(\hat X_{1, n+h}^{\lambda}, \ldots, \hat X_{q, n+h}^{\lambda})' ,
\] 
where $L$ is known and $\hat X_{j, n+h}^{\lambda}$, for every $j = 1, \ldots, q$, are predicted using time series ARIMA approach.

Such aggregation might induce a smaller loss in forecast accuracy of factors $F$ when we fail to accurately define a direct model. First, forming the extracted factors with weights of a similar size, such forecast aggregation is equivalent to \textit{bagging} (bootstrap aggregation as in \cite{Breiman1996}). Second, it is possible, that the data generating process of $F_j$ might be from some family of complex, long memory processes. Therefore, the aggregation of forecasts of simpler models of its components introduces some degree of freedom to make inaccurate estimations of components models while still generating more accurate final predictions. Indeed, \cite{GRANGER1980227} has shown that the aggregation of a low-order AR/ARMA processes in particular cases may produce a process with more complex dynamics. Extending these ideas to the aggregation of the forecasted series allows recovering such complex dynamics in the final forecasts of the original factors. We demonstrate the possible gains from such approach in this paper by comparing the resulting forecasts with both direct factor forecasts (see, e.g., Table \ref{gfcf:targeted}) and Kalman filter forecasts, used with the dynamic factors (see, e.g., Table \ref{gfcf:main2} and similar).

Noteworthy, when the variables $X_j$ are orthogonal, due to the properties of Principal Components, the transformation would reduce to the base LASSO performance, since the total amount of information would remain unchanged.

\subsection{Tailoring the rotation of Principal Components}
\label{sparsePCA}
Having introduced the general idea of the transformation, it is possible to extend the method by adding a certain degree of supervision to the standard Principal Components approach. The classic transformation is convenient in such a way that the rotation matrix can be tailored against the specific modelled variable by modifying the scales or the angles of the components.

Scale modification might help the LASSO distinguishing the most important signals since the components with larger-scale will enter the solution path earlier than the other, smaller components. This idea is discussed by \cite{RePEc:lie:wpaper:13}, who observes 
significant improvement when using Weighted PCA or Generalized PCA for the extraction of factors when nowcasting Lithuanian GDP.

Alternatively, in this section, we discuss the idea of angle modification by sparsifying the rotation matrix, e.g., through Sparse PCA (\cite{10.2307/27594179}). The idea here is that we may achieve a more accurate factor representation by losing the orthogonality of the transformation. The main motivation is that, even though the data matrix $X$ is preselected by the LASSO as a matrix, containing mainly significant variables, it is not clear that by rotating the variables to the latent space, all of them will be significant there. In other words, if there are two strongly correlated variables preselected by LASSO as significant, both having roughly the same estimated weights (possibly with different signs) in the loading matrix, it is likely that losing one of the two dimensions might not change the predictive power of the resulting factor estimate.

Further, some of the variables denoted in what follows as $Z$, where $Z \subset \{X_{j}: j = 1, \ldots, q \}$, might be orthogonal to all other preselected variables, meaning that the principal component solution does not extract the correct factor of it from the latent space. That is, in the latent space, ideally, they would form a direction, where the coordinate vector would have zeros for all other variables. However, in standard principal components that is mostly not the case. Every extracted factor is a linear combination of all the variables used, even if the weights are close to zero, it's unlikely for them to be exactly zero. Let us assume that the model matrix $X^{\lambda}$ is reordered such, that the block of  variables $Z \in \mathbb{R}^{q_0 \times n}, q_0 < q$, is able to capture significant explanatory signals to our modelling problem, orthogonal to the remaining block $\bar X$, the LASSO will try to extract as much information as possible from those variables. Since the extracted factor matrix has the following structure:

\[
F = L' X^\lambda = L' \cdot (Z', \bar X')'
\]
the $j-$th factor component will have the following structure:
\[
 f_j = [L']_j \cdot (Z', \bar X')' = [\Lambda_{q \times q_0}, \Phi_{q \times (q_0-q)}]_j\cdot (Z', \bar X')'. 
\]

Therefore LASSO, while trying to reconstruct the signal from $Z$, will include too many factors $f_{j}$ to the final solution, since all of them will have some information from $Z$. Some of those factors would not be included if part of the rotation matrix would have zeros  -- the $\Phi$ block in this example. Let's assume that $\mathcal{G} \subset \{1, \ldots, q\}$ is a set of indices denoting factors $f_{j}$, which have been selected as significant by the LASSO in the final solution only because of non-zero loading weights in $\Phi_{j}$. Then, with every additional $f_{r}, ~r \in \mathcal{G}$,  included we will add some noise in the scale of $\Phi_r \bar X'$ to the data. And the more such factors are selected, the closer the LASSO-PC solution is to the usual Relaxed LASSO solution. 

From this discussion, we can see the benefit of adding a step to the LASSO-PC procedure. One way is to modify the loading matrix $L$ to introduce some sparseness to it (e.g., by using Sparse PCA (SPCA), \cite{10.2307/27594179}). An alternative way could be including the preselected data matrix $X^{\lambda}$  together with the extracted factors $F$ and model them together as $\bar F = (F', X')'$ 

using the LASSO. While it may seem that in this way no new information is added, it potentially gives the LASSO additional degrees of freedom to select the necessary combination of variables, in essence recovering the best fitting loading matrix.
Thus, the Sparse PCA, base LASSO or the LASSO-PC would potentially become specific cases, depending on the estimated values of $\bar F_j$ coefficients. 

\section{Preliminaries: data preparation}
\label{duomenys}

In this paper, we consider the four main components of the GDP by the expenditure approach: Gross Fixed Capital Formation (GFCF), Private Final Consumption Expenditure (PFCE), Imports and Exports of goods and services, all of which are quarterly and seasonally adjusted by the source. 

The monthly data used as explanatory variables are various openly available indicators from the databases of St. Louis Bank of Federal Reserves (FRED) and Eurostat from 1990 to 2019, with up to 2400 various macroeconomic time series used in total. Each time series used in the modelling were either seasonally adjusted by the source or by using the X13-ARIMA-SEATS procedure for seasonal adjustment. Avoiding the problem of spurious regression, every time series were transformed to stationary and appropriately adjusted for normality and against additive outliers. Dealing with a large number of variables and relying on automatic algorithms for variable selection without any prior expectations, strict rules and conservative approach for data preparation allows ruling out as much as possible noise  (see Appendix \ref{appendix:data} for specific heuristics and tests applied).

Comparing the forecasting performances of different models in a realistic setting, during the pseudo-real-time experiments, the results of which are presented in Section \ref{chapter:PR}, we reconstructed the pseudo-real-time dataset for every iteration of the exercise by adjusting the amount of available data by the appropriate release lag for each monthly indicator (see Table \ref{table:PR} for detailed illustration). During a full quarter, at least three updates on the dataset are possible for every different month of the quarter. However, in this paper, the results presented are of the last month of the full quarter. Since we are also interested in nowcasting performance, analyzing results of 3rd month helps to separate it from forecasting due to available indicators with low publication lag.  Otherwise, we would just be comparing the predictive performance of the ARIMA models, used for the individual predictions of the selected monthly indicators, which is already inspected using 1- and 2-quarter forecasts.

\makeatletter
\def\highlight#1{%
	\fboxrule2pt %
	\hsize=\dimexpr\hsize-2\fboxrule-2\fboxsep\relax
	#1%
	\@endpbox\unskip\setbox0\lastbox\bgroup
	\fboxrule2pt %
	\fcolorbox{red}{lightgray}{\box0}\hfill}

\begin{table}
	
	\scriptsize
	
	\caption{\label{table:PR} The timing and data availability framework for the pseudo-real-time forecasting exercise}
	\begin{tabular}{clcccccccccccccc }
		
		\hline
		
		\textbf{Target}& GFCF&\multicolumn{3}{|c|}{\cellcolor[HTML]{C0C0C0} Backcast, Q1} &\multicolumn{3}{|c|}{\cellcolor[HTML]{DCDCDC} Nowcast, Q2}&\multicolumn{3}{|c|}{ \cellcolor[HTML]{DCDCDC} Forecast, Q3}&\multicolumn{3}{|c|}{\cellcolor[HTML]{DCDCDC} Forecast, Q4}& Quarter \\
		\hline 
		&&~~1&~~2&~~3&~~4&~~5& \boxit{0.182in} ~~6&~~7&~~8&~~9& ~~10&~~11&~~12 & Month\\
		\hline
		\multirow{5}{*}{\textbf{Group}}& Industrial Production&\cellcolor[HTML]{C0C0C0}&\cellcolor[HTML]{C0C0C0}&\cellcolor[HTML]{C0C0C0}& \cellcolor[HTML]{C0C0C0} &\cellcolor[HTML]{DCDCDC}  &\cellcolor[HTML]{DCDCDC}&\cellcolor[HTML]{DCDCDC}&\cellcolor[HTML]{DCDCDC}&\cellcolor[HTML]{DCDCDC}&\cellcolor[HTML]{DCDCDC}&\cellcolor[HTML]{DCDCDC}&\cellcolor[HTML]{DCDCDC}\\
		&Surveys&\cellcolor[HTML]{C0C0C0}&\cellcolor[HTML]{C0C0C0}&\cellcolor[HTML]{C0C0C0}& \cellcolor[HTML]{C0C0C0} &\cellcolor[HTML]{C0C0C0}  &\cellcolor[HTML]{DCDCDC}&\cellcolor[HTML]{DCDCDC}&\cellcolor[HTML]{DCDCDC}&\cellcolor[HTML]{DCDCDC}&\cellcolor[HTML]{DCDCDC}&\cellcolor[HTML]{DCDCDC}&\cellcolor[HTML]{DCDCDC}\\
		&Interest Rates&\cellcolor[HTML]{C0C0C0}&\cellcolor[HTML]{C0C0C0}&\cellcolor[HTML]{C0C0C0}& \cellcolor[HTML]{C0C0C0} &\cellcolor[HTML]{C0C0C0}  &\cellcolor[HTML]{C0C0C0}&\cellcolor[HTML]{DCDCDC}&\cellcolor[HTML]{DCDCDC}&\cellcolor[HTML]{DCDCDC}&\cellcolor[HTML]{DCDCDC}&\cellcolor[HTML]{DCDCDC}&\cellcolor[HTML]{DCDCDC}\\
		&Stock Market Indexes&\cellcolor[HTML]{C0C0C0}&\cellcolor[HTML]{C0C0C0}&\cellcolor[HTML]{C0C0C0}& \cellcolor[HTML]{C0C0C0} &\cellcolor[HTML]{C0C0C0}  &\cellcolor[HTML]{C0C0C0}&\cellcolor[HTML]{DCDCDC}&\cellcolor[HTML]{DCDCDC}&\cellcolor[HTML]{DCDCDC}&\cellcolor[HTML]{DCDCDC}&\cellcolor[HTML]{DCDCDC}&\cellcolor[HTML]{DCDCDC}\\
		&Trade&\cellcolor[HTML]{C0C0C0}&\cellcolor[HTML]{C0C0C0}&\cellcolor[HTML]{C0C0C0}& \cellcolor[HTML]{DCDCDC} &\cellcolor[HTML]{DCDCDC}  &\cellcolor[HTML]{DCDCDC}&\cellcolor[HTML]{DCDCDC}&\cellcolor[HTML]{DCDCDC}&\cellcolor[HTML]{DCDCDC}&\cellcolor[HTML]{DCDCDC}&\cellcolor[HTML]{DCDCDC}&\cellcolor[HTML]{DCDCDC}\\
		\hline
		\\
		\textbf{Legend:}&\multicolumn{1}{r}{Available data:}&\cellcolor[HTML]{C0C0C0}&\multicolumn{4}{r}{Month of forecast:}&\boxit{0.182in} ~~~&\multicolumn{4}{r}{Forecasted data:}&\cellcolor[HTML]{DCDCDC}&\\
		
	\end{tabular}

\end{table}

The monthly variables used in the dataset were aggregated to quarterly by averaging, with up to four quarterly lags included, and the ragged edges of the dataset were filled by using the ARIMA time series methods.  

\section{Pseudo-real-time forecasting experiments}
\label{chapter:PR}
This section presents the results of pseudo-real-time forecasting exercise over 2005Q1 -- 2019Q1. The main goal is to gauge the real-world performance of the sparse methods in a high-dimensional environment. As discussed in Section \ref{duomenys}, we consider as much macroeconomic data available as possible (potentially including some noise variables), and are interested in 4 specific forecast horizons: one backcast, one nowcast and two predictions of 1- and 2-quarters ahead. We focus on the main four components of the GDP by expenditure approach (GFCF, PFCE, Imports and Exports of goods and services) for five economies: the US, France, Italy, Spain and Germany. We present the main results and important discussion points using the US data, while the European countries are presented as a robustness check to inspect how well do the results translate between the different datasets. 
 
In sum, the expenditure components reflect the main drivers of the economy: the domestic demand, mainly consisting of private consumption, investments and imports, and the foreign demand defined by exports. The international trade reflects the openness of the economy and drives international competitiveness that forces the search for new innovative and advanced solutions.

Out of these four variables, the most difficult to accurately predict is the GFCF composed of investments in many different industries. The investment spending deemed to be forward-looking and optimistic, expanding rapidly during an economic boom when investors expect that the future will require the extensive productive capacity, and falling fast when such expectations evaporate. Therefore, it is the most volatile of the four variables. For private consumption, imports and exports, there are ``hard'' coinciding monthly indicators available, allowing for an easier nowcast. However, such variables are absent for the GFCF making good nowcasts a more challenging task. Therefore, our primary focus in this section is forecasting the GFCF presented in greater detail. The forecasting results of the three remaining variables are then briefly discussed.

 The analysis aims to inspect the model robustness to the changing nature of the economy, testing if the models can reflect and respond to possible structural changes, as the economy evolves and matures. Therefore, we expect maximum flexibility by using all of the available data. Further, we seek to test the signal recovery and identification of the sparse structures. On the one hand, we allow the inclusion of possible noise variables in the datasets, for example, various indicators of foreign economies. On the other hand, this inclusion provides the possibility for the sparse models to recover significant signals from large trade partners and dependent parties without specific offsets imposed by the analyst. This exercise may present itself a perfect tool for additional variable consideration to their proprietary datasets, likely enhancing any existing models with otherwise omitted stories, as we see later in this paper.

\subsection{Setting up the experiment}

For every modelled component of the GDP, we consider the following models and their corresponding notations: the Square-Root LASSO (in the tables denoted as SqL), Relaxed LASSO ({RL}), Adaptive LASSO ({AdL}), classic LASSO ({L}), the L0Lq best subset selection models (L0, L0L1, L0L2) and a proposed combination LASSO-PC, where the data is preselected by a specific variant of the LASSO, i.e., Adaptive LASSO-PC ({AdP}), L0-PC ({L0P}), L0L1-PC ({L0L1P}), L0L2-PC ({L0L2P}), Square-Root LASSO-PC ({SqP}) or the LASSO-PC ({LP}), and similar, as described in Section \ref{PCA}. In what follows, the letter ``P'' is always appended to any model abbreviation where the PCA transformation is applied. 

Seeking to inspect the gains brought solely by performing the rotation of the data to its principal components, we analyse the alternative cases for AdP models fixing the preselected variables without the rotation to the principal components. In a way, it becomes a mix between Relaxed LASSO and Adaptive LASSO, denoted as {AdRL} in the tables.  Additionally, in cases when there were significant gains in forecasting accuracy when using both the rotated and original data, we interpret such cases as a cross between the LASSO and LASSO-PC methods denoting them as LASSO-PC-X ({LPX} in the tables). 

Each LASSO model is estimated using the cross-validated hyperparameters unless specified differently, where each hyperparameter is chosen to maximize the out-of-sample accuracy (\cite{chetverikov2016crossvalidated}). Besides, the weights for Adaptive LASSO are chosen using Ridge estimates (as discussed in Section \ref{adaLasso}) with $\gamma \in \{0.001$, $0.01$, $0.1$, $0.2$, $0.3$, $0.5$, $1$, $1.5$, $2\} $ selected by cross-validation (\cite{doi:10.1198/016214506000000735}). The idea behind adding values of $\gamma$ very close to 0 is to allow the Adaptive LASSO optimize under weights that can be close to the LASSO. 

As benchmarks for the analysed models we use ARMA$(p,q)$ models, where the orders $(p, q)$ are selected to minimize the corrected \textit{Akaike} information criterion (AICc, \cite{JSSv027i03}) during each quarter of the exercise. Additionally, we consider a few variants of Factor models as good \textit{dense} structure alternatives based on their known performance in the literature. Namely, we consider the static Diffusion-Index models ({DI}, \cite{RePEc:bes:jnlbes:v:20:y:2002:i:2:p:147-62})   and Dynamic Factor models ({DF}, \cite{GIANNONE2008665},  \cite{reichlin11}). For the dynamic factor estimation we use the two-step Kalman Filter realization with quarterly aggregation scheme (\cite{Angelini2011}).
 Additionally, for the DF models, we consider the targeted predictor approach of \cite{Bai2008304} (denoted in the tables as BNDF), i.e., instead of full model matrix, we preselect the candidate variables using the LASSO. Lastly, for more clarity and better comparison, during each iteration of the exercise, if some variant of LASSO was used for initial variable selection, e.g., for LASSO-PC, Relaxed LASSO and BNDF, the same variables will be used by these models. This was done to eliminate the uncertainty from randomness in variable selection. For all factor models, the number of significant factors was estimated based on the information criterion, defined by \cite{doi:10.1111/1468-0262.00273}. Additionally, for the DF models, the number of quarterly factor lags is fixed to 2, while the number of shocks is estimated by information criterion, following the \cite{10.2307/27638906}.

Besides, we present the results of models, where, instead of the cross-validated hyperparameters, we use the fixed number of selected variables that provide a more parsimonious result. In those cases the results are presented with an additional number next to their abbreviation, e.g., {RL15} in Table \ref{gfcf:RLvsPP}, indicating that the model is selecting top 15 significant variables during the exercise following \cite{RollingWindow}, while \cite{Bai2008304} claim that practical choice of top predictors should not exceed 30. It is worth inspecting such results since the optimally selected hyperparameter can result in a model of a too dense structure. In this case, the LASSO wins a small amount of validation accuracy but brings an additional forecast uncertainty with each of the extra variable included.

During the forecasting exercise, the subsets of variables are reselected for each analyzed model and every quarter when the new pseudo-vintage of data is formed. Instead of tailoring the set of possible variables to match the predicted variable, we allow methods themselves to select the significant covariates. Likely, some indicators that drove the growth of particular markets in the past are not significant at a later stage. Hence, these past drivers can be replaced on time with new indicators reflecting new products and services. For this reason, we study two different approaches to forming the training sample: an expanding window approach and a rolling window approach. 

For the expanding window experiments, we use the sample from 1992Q1 to 2004Q4, the window is expanded by adding one additional quarter during every iteration. The size of the window was chosen to be moderate so that there would be an appropriate amount of available historical monthly indicators, but sufficient for the methods to select a large number of significant variables if needed. Note, that under such an approach it is likely that some of the variables are capturing the historical dynamics of the modelled series early in the training sample rather than of the more recent events. Therefore, with a rolling window approach, we add some control over possible structural breaks (\cite{RollingWindow}). We chose a 13-year rolling window to capture at least one full business cycle (see, e.g., Economic Cycle Research Institute (ECRI), \cite{CycleDuration}): starting from 1992Q1 to 2004Q4, the window was rolled by adding a quarter to both the start and the end of the period during every iteration. 

To indicate how well did the models forecast, we present the ratio of the RMSE of the LASSO models to the list of benchmark models, in addition to RMSE and pairwise Diebold-Mariano (DM, \cite{Diebold}) tests on the forecast errors for the expanding window exercise, and pairwise \cite{Giacomini} (GW) tests for the rolling window exercise.

\subsection{Expanding window: US Gross Fixed Capital Formation}

\label{expW:gfcf}

\subsubsection{Main findings}
In this section we present the results of forecasting GFCF during the period of 2005Q1-2019Q1. The results of RMSE of the forecasted values are presented in Table \ref{gfcf:main1} for all models and for all 4 forecast horizons. The total period results were divided into three periods (2005Q1-2007Q4, 2008Q1-2014Q4, 2015Q1-2019Q1) aiming to compare the models during different stages of the economy: during the stable growth of 2005Q1-2007Q4, the crisis and the recovery period of 2008Q1-2014Q4, and the stable growth of 2015Q1-2019Q1. Dividing up the total results is also useful in inspecting whether there are models, dominating every other competitor in the ``horse race''. Also, in Table \ref{gfcf:main2} the results of Relative RMSE are compared with the performance of the benchmark ARMA, BNDF, DF and SW models. 

\setlongtables{\scriptsize
	\setlength\tabcolsep{4.5pt}
	\begin{longtable}{lcccc|cccc|cccc|ccccccccccccccc}\caption{RMSE of models forecasts during pseudo-real experiments for GFCF. The labels B, N, Q1 and Q2 denotes the forecast horizons -- Backcast, Nowcast, Forecast-1Q and Forecast-2Q, respectively. The bolded value is the smallest one for every column; the blocks correspond to different time periods, while the last block shows performance over the full 2005-2019 period. \label{gfcf:main1}} \tabularnewline \hline\hline
		\bfseries \multirow{2}{*}{Model}&\multicolumn{4}{c}{\bfseries 2005Q1-2007Q4}&\multicolumn{4}{c}{\bfseries 2008Q1-2014Q4}&\multicolumn{4}{c}{\bfseries 2015Q1-2019Q1}&\multicolumn{4}{c}{\bfseries 2005Q1-2019Q1 \rule{0pt}{2.5ex} } \tabularnewline*
		&\bfseries B&\bfseries N&\bfseries Q1&\bfseries Q2&\bfseries B&\bfseries N&\bfseries Q1&\bfseries Q2&\bfseries B&\bfseries N&\bfseries Q1&\bfseries Q2&\bfseries B&\bfseries N&\bfseries Q1&\bfseries Q2 \rule{0pt}{2.5ex} \tabularnewline*
		\hline
		\endhead
		\hline 
		\endfoot
		L \rule{0pt}{2.5ex}&0.70&0.92&1.14&1.27&0.80&1.72&2.48&2.74&0.58&0.82&0.90&0.85&\textcolor{black}{0.65}&\textcolor{black}{1.09}&\textcolor{black}{1.45}&\textcolor{black}{1.56} \\ 
		LP &0.53&\textbf{0.84}&1.10&1.23&0.53&1.49&\textbf{2.27}&2.75&0.49&0.86&0.91&0.84&\textcolor{black}{0.50}&\textcolor{black}{1.01}&\textcolor{black}{\textbf{1.37}}&\textcolor{black}{1.56} \\ 
		LPX&0.45&0.85&1.06&1.22&0.50&1.47&2.31&2.73&0.49&0.88&0.93&0.84&\textcolor{black}{0.47}&\textcolor{black}{1.01}&\textcolor{black}{1.38}&\textcolor{black}{1.56} \\ 
		RL&0.52&0.84&1.05&1.19&0.52&\textbf{1.44}&2.32&2.69&0.45&0.92&0.97&0.87&\textcolor{black}{0.47}&\textcolor{black}{1.03}&\textcolor{black}{1.41}&\textcolor{black}{1.55} \\ 
		SqL&0.70&0.89&1.11&1.25&0.71&1.67&2.49&2.76&0.58&0.82&0.88&0.84&\textcolor{black}{0.62}&\textcolor{black}{1.06}&\textcolor{black}{1.44}&\textcolor{black}{1.57} \\ 
		SqP&0.52&0.89&1.01&1.16&0.44&1.60&2.44&2.68&0.51&0.83&0.86&0.83&\textcolor{black}{0.48}&\textcolor{black}{1.05}&\textcolor{black}{1.39}&\textcolor{black}{1.52} \\ 
		AdL&0.07&1.04&\textbf{0.98}&1.35&0.19&1.85&2.56&2.88&0.22&0.87&0.95&0.92&\textcolor{black}{0.18}&\textcolor{black}{1.18}&\textcolor{black}{1.47}&\textcolor{black}{1.65} \\ 
		AdP&0.07&0.99&1.01&1.30&0.12&1.85&2.47&2.87&0.21&0.92&0.96&0.91&\textcolor{black}{0.16}&\textcolor{black}{1.17}&\textcolor{black}{1.44}&\textcolor{black}{1.64} \\ 
		AdRL&\textbf{0.04}&1.01&1.00&1.35&\textbf{0.11}&1.86&2.45&2.82&\textbf{0.15}&0.92&0.96&0.91&\textcolor{black}{\textbf{0.12}}&\textcolor{black}{1.20}&\textcolor{black}{1.42}&\textcolor{black}{1.62} \\ 
		L0&0.58&1.09&1.10&1.28&1.12&2.02&2.52&2.58&0.56&0.88&1.04&0.83&\textcolor{black}{0.74}&\textcolor{black}{1.23}&\textcolor{black}{1.49}&\textcolor{black}{1.50} \\ 
		L0p&0.57&1.13&1.09&1.31&1.07&2.07&2.59&2.60&0.56&0.87&1.05&0.83&\textcolor{black}{0.72}&\textcolor{black}{1.25}&\textcolor{black}{1.52}&\textcolor{black}{1.51} \\ 
		L0L1&0.82&0.92&1.21&1.37&0.95&1.67&2.34&2.51&0.61&0.84&0.90&\textbf{0.82}&\textcolor{black}{0.74}&\textcolor{black}{1.10}&\textcolor{black}{1.42}&\textcolor{black}{\textbf{1.49}} \\ 
		L0L1p&0.80&0.94&1.23&1.38&0.92&1.70&2.42&2.54&0.59&0.86&0.93&0.84&\textcolor{black}{0.73}&\textcolor{black}{1.13}&\textcolor{black}{1.47}&\textcolor{black}{1.51} \\ 
		L0L2&0.82&0.95&1.22&1.36&1.05&1.76&2.32&\textbf{2.50}&0.60&\textbf{0.74}&0.87&0.84&\textcolor{black}{0.77}&\textcolor{black}{1.11}&\textcolor{black}{1.40}&\textcolor{black}{1.49} \\ 
		L0L2p&0.80&0.95&1.24&1.39&0.99&1.87&2.48&2.50&0.56&0.77&0.90&0.87&\textcolor{black}{0.73}&\textcolor{black}{1.16}&\textcolor{black}{1.48}&\textcolor{black}{1.50} \\ 
		\hline
		ARMA \rule{0pt}{2.5ex} &0.86&0.84&1.04&\textbf{1.15}&1.98&2.56&3.16&3.44&0.92&0.99&1.02&1.07&\textcolor{black}{1.24}&\textcolor{black}{1.48}&\textcolor{black}{1.77}&\textcolor{black}{1.91} \\ 
		BNDF&0.78&0.92&1.16&1.36&1.09&1.46&2.36&3.03&0.75&0.83&0.90&0.85&\textcolor{black}{0.81}&\textcolor{black}{\textbf{1.00}}&\textcolor{black}{1.39}&\textcolor{black}{1.70} \\ 
		DF&1.12&1.18&1.42&1.57&1.16&1.49&2.30&2.82&0.87&1.14&1.14&1.04&\textcolor{black}{0.98}&\textcolor{black}{1.18}&\textcolor{black}{1.50}&\textcolor{black}{1.71} \\ 
		SW&1.40&1.45&1.57&1.69&1.47&1.96&3.39&3.98&0.89&0.90&\textbf{0.83}&0.85&\textcolor{black}{1.14}&\textcolor{black}{1.30}&\textcolor{black}{1.86}&\textcolor{black}{2.13} \\ 
		
\end{longtable}}

\setlongtables{\scriptsize
	\begin{longtable}{lclllclllclllclll}\caption{US: GFCF. Relative RMSE of model forecasts during the expanding window pseudo-real-time experiments against four benchmarks: ARMA, BNDF, DF and SW models. The labels N, Q1 and Q2 denotes the forecast horizons -- Nowcast, Forecast-1Q and Forecast-2Q, respectively. The bolded values correspond to the greatest accuracy within the forecast horizon, while the asterix denotes significant performance improvement as suggested by DM test with 5\% significance. \label{gfcf:main2}} \tabularnewline
		\hline\hline
		\multicolumn{1}{c}{\bfseries Model}&\multicolumn{1}{c}{\bfseries \rule{0pt}{2.5ex}  \vspace{0.02 in}}&\multicolumn{3}{c}{\bfseries ARMA}&\multicolumn{1}{c}{\bfseries }&\multicolumn{3}{c}{\bfseries BNDF}&\multicolumn{1}{c}{\bfseries }&\multicolumn{3}{c}{\bfseries DF}&\multicolumn{1}{c}{\bfseries }&\multicolumn{3}{c}{\bfseries SW}\tabularnewline
		\cline{3-5} \cline{7-9} \cline{11-13} \cline{15-17}
		\multicolumn{1}{c}{}&\multicolumn{1}{c}{\rule{0pt}{2.5ex}  \vspace{0.02 in}}&\multicolumn{1}{c}{\textbf{N}}&\multicolumn{1}{c}{\textbf{Q1}}&\multicolumn{1}{c}{\textbf{Q2}}&\multicolumn{1}{c}{}&\multicolumn{1}{c}{\textbf{N}}&\multicolumn{1}{c}{\textbf{Q1}}&\multicolumn{1}{c}{\textbf{Q2}}&\multicolumn{1}{c}{}&\multicolumn{1}{c}{\textbf{N}}&\multicolumn{1}{c}{\textbf{Q1}}&\multicolumn{1}{c}{\textbf{Q2}}&\multicolumn{1}{c}{}&\multicolumn{1}{c}{\textbf{N}}&\multicolumn{1}{c}{\textbf{Q1}}&\multicolumn{1}{c}{\textbf{Q2}}\tabularnewline
		\hline
		\endfirsthead\caption[]{\em (continued)} \tabularnewline
		\hline
		\multicolumn{1}{c}{\bfseries Model}&\multicolumn{1}{c}{\bfseries\rule{0pt}{2.5ex}  \vspace{0.02 in} }&\multicolumn{3}{c}{\bfseries ARMA}&\multicolumn{1}{c}{\bfseries }&\multicolumn{3}{c}{\bfseries BNDF}&\multicolumn{1}{c}{\bfseries }&\multicolumn{3}{c}{\bfseries DF}&\multicolumn{1}{c}{\bfseries }&\multicolumn{3}{c}{\bfseries SW}\tabularnewline
		\cline{3-5} \cline{7-9} \cline{11-13} \cline{15-17}
		\multicolumn{1}{c}{\rule{0pt}{2.5ex}  \vspace{0.02 in}}&\multicolumn{1}{c}{}&\multicolumn{1}{c}{\textbf{N}}&\multicolumn{1}{c}{\textbf{Q1}}&\multicolumn{1}{c}{\textbf{Q2}}&\multicolumn{1}{c}{}&\multicolumn{1}{c}{\textbf{N}}&\multicolumn{1}{c}{\textbf{Q1}}&\multicolumn{1}{c}{\textbf{Q2}}&\multicolumn{1}{c}{}&\multicolumn{1}{c}{\textbf{N}}&\multicolumn{1}{c}{\textbf{Q1}}&\multicolumn{1}{c}{\textbf{Q2}}&\multicolumn{1}{c}{}&\multicolumn{1}{c}{\textbf{N}}&\multicolumn{1}{c}{\textbf{Q1}}&\multicolumn{1}{c}{\textbf{Q2}}\tabularnewline
		\hline
		\endhead
		\hline
		\endfoot
		
		AdL&\rule{0pt}{2.5ex} &0.79&0.83&0.87&&1.18&1.05&0.97&&0.99&0.98&0.97&&0.90&0.79&0.78\tabularnewline
		AdP&&0.79&0.81&0.86&&1.17&1.03&0.96&&0.99&0.96*&0.96&&0.90&0.77&0.77\tabularnewline
		AdRL&&0.81&0.80&0.85&&1.21&1.02&0.96&&1.02&0.95*&0.95&&0.92&0.76&0.76\tabularnewline
		L&&0.73*&0.82&0.82&&1.09&1.04&0.92&&0.92&0.97*&0.92*&&0.83&0.78&0.73\tabularnewline
		L0&&0.83&0.84&0.79&&1.23&1.07&0.89&&1.04&0.99&0.88&&0.94&0.80&0.70\tabularnewline
		L0L1&&0.74*&0.80&0.78&&1.11&1.02&0.88&&0.93&0.95&0.87&&0.84&0.76&0.70\tabularnewline
		L0L1p&&0.76&0.83&0.79&&1.13&1.05&0.89&&0.95&0.98&0.88&&0.86&0.79&0.71\tabularnewline
		L0L2&&0.75*&0.79&\textbf{0.78}&&1.11&1.01&\textbf{0.88}&&0.93&0.94&\textbf{0.87}&&0.85&0.75&\textbf{0.70}\tabularnewline
		L0L2p&&0.79&0.83&0.79&&1.17&1.06&0.89&&0.98&0.98&0.88&&0.89&0.79&0.70\tabularnewline
		L0p&&0.85&0.86&0.79&&1.26&1.09&0.89&&1.06&1.01&0.89&&0.96&0.81&0.71\tabularnewline
		LP&&\textbf{0.68*}&\textbf{0.77}&0.82&&\textbf{1.02}&\textbf{0.98}&0.92&&\textbf{0.86*}&\textbf{0.91*}&0.92*&&\textbf{0.78*}&\textbf{0.73}&0.73*\tabularnewline
		LPX&&0.68*&0.78&0.82&&1.02&0.99&0.92&&0.86&0.92*&0.91*&&0.78*&0.74&0.73*\tabularnewline
		RL&&0.70&0.79&0.81&&1.04&1.01&0.91&&0.87&0.94&0.91*&&0.79&0.76&0.73\tabularnewline
		SqL&&0.72*&0.81&0.82&&1.07&1.03&0.92&&0.90*&0.96*&0.92*&&0.81*&0.77&0.74\tabularnewline
		SqP&&0.71&0.79&0.80&&1.06&1.00&0.90&&0.89&0.93*&0.89*&&0.81&0.75&0.71*\tabularnewline
		\hline
\end{longtable}}

The results reveal that when forecasting the GFCF most of the sparse models provide a rather similar forecasting performance, including the BNDF with the LASSO based preselection of the set of variables. The dense benchmark models (SW and DF) seem to fall behind, mostly due to underpredicted shock period of 2008Q1-2014Q4. Finally, the ARMA benchmark models show the least overall accurate predictions, except for the period of 2005Q1-2007Q4 when forecasting 2-quarters ahead, where they are able to outperform every other model by a small margin. Such a result is consistent with the literature: e.g., \cite{RePEc:ecb:ecbwps:20060680} highlights the fact that during relatively steady growths (the authors analysed the Great Moderation period in particular, where a sizeable decline in volatility of output and price measures was observed) even sophisticated models can fail to outperform simple AR models. Therefore, analysis of the recession period of 2008Q1-2014Q4 is the most interesting one, since then we are comparing the performance of models during a unique event with no historical precedent. Overall, these results further emphasize the value of additional monthly data included in the modelling, especially during the more volatile periods of 2005Q1-2014Q4. 

Besides, the RL and SqL models are able to increase the predictive performance of the regular LASSO just as expected. However, the gains for Adaptive and L0Lq variants are not that evident:  all of the L0Lq models were able to improve the 2-quarter forecast performance, with the L0L2 having the most accurate 2-quarter forecasts, without significant improvements for the nowcasts or 1-quarter forecasts. The comparison between these groups can be interesting from the variable selection perspective -- in most cases throughout the experiment, the AdL models were selecting the median of 55 variables, while the RL and SqL ranged between 23-26 variables, and L0Lq models between 2-9 (L0 was the most sparse, with the median of 2 variables selected). Evidently, these numbers translate well into the overall accuracy results: too large amount of variables selected likely introduced additional level of noise into the forecasts, while the too sparse models were likely able to only capture the long term dynamics, but lacking information for nowcasts and 1-quarter forecasts -- the differences are most evident during the 2008Q1-2014Q4 period.

 Moreover, the results show that the usage of PCA in the estimation can additionally improve the predictive performance of the models: in most of the cases proposed modifications are able to outperform their Relaxed counterparts (the LP vs RL, SqP vs SqL, AdP vs AdL), even though by a small  margin. However, for the L0Lq models, likely due to the small amount of variables selected the transformation gains were not evident. 

Additionally, in Table \ref{gfcf:main2} we mark the significance in forecast gains based on the DM test with 5\% significance. First, it should be noted that all of the LASSO models are able to outperform ARMA models when nowcasting. Further, the LP and SqL are able to significantly outperform DF and SW models in nowcasting, while in 1- and 2-quarter forecasting most variants of RL and SqL were able to significantly outperform DF and SW models. Overall the most accurate nowcasts are generated by the BNDF model\footnote{It is worth highlighting the high computational complexity of the DF models in comparison to sparse methods, making the estimation very resource demanding when working with more than 2000 variables as in the case of this paper, yet still showing similar (or weaker) forecasting performance to various sparse methods.}.

\setlongtables{\scriptsize
	\begin{longtable}{lcc|cc|cc|ccccccccccc}\caption{ RMSE of models forecasts during pseudo-real experiments for GFCF. The bolded value is the smallest one for every row within the same group by number of variables selected. For every block the last line denotes the total error for the period 2005-2019. \label{gfcf:RLvsPP}}\tabularnewline \hline
		\multicolumn{1}{l}{\bfseries \rule{0pt}{2.5ex}  \vspace{0.02 in} }&\multicolumn{1}{c}{\bfseries LP10}&\multicolumn{1}{c}{\bfseries RL10}&\multicolumn{1}{c}{\bfseries LP15}&\multicolumn{1}{c}{\bfseries RL15}&\multicolumn{1}{c}{\bfseries LP20}&\multicolumn{1}{c}{\bfseries RL20}&\multicolumn{1}{c}{\bfseries LP30}&\multicolumn{1}{c}{\bfseries RL30}\tabularnewline*
		\hline
		\endfirsthead
		\caption{ (Continued) RMSE of models forecasts during pseudo-real experiments for GFCF. The bolded value is the smallest one for every row within the same group by number of variables selected. For every block the last line denotes the total error for the period 2005-2019.}
		
		\endhead
		\hline
		\endfoot
		\rule{0pt}{2.5ex} 
		{\bfseries Backcast}&&&&&&&&\tabularnewline*
		\rule{0pt}{2.5ex} 
		~2005Q1-2007Q4&\textbf{0.58}&0.64&0.50&\textbf{0.47}&0.39&\textbf{0.34}&0.29&\textbf{0.26}\tabularnewline*
		~~2008Q1-2014Q4&\textbf{0.92}&0.94&\textbf{0.52}&0.55&0.58&\textbf{0.50}&\textbf{0.41}&0.42\tabularnewline*
		~~2015Q1-2019Q1&\textbf{0.70}&0.70&\textbf{0.61}&0.62&0.62&\textbf{0.57}&0.52&\textbf{0.47}\tabularnewline*
		\vspace{0.02 in}
		~\textcolor{black}{2005Q1-2019Q1}&\textcolor{black}{\textbf{0.70}}&\textcolor{black}{0.71}&\textcolor{black}{\textbf{0.55}}&\textcolor{black}{0.56}&\textcolor{black}{0.54}&\textcolor{black}{\textbf{0.50}}&\textcolor{black}{0.44}&\textcolor{black}{\textbf{0.41}}\tabularnewline
		\hline
		\rule{0pt}{2.5ex} 
		{\bfseries Nowcast}&&&&&&&&\tabularnewline*
		\rule{0pt}{2.5ex} 
		~2005Q1-2007Q4&0.83&\textbf{0.78}&0.89&\textbf{0.81}&0.87&\textbf{0.84}&\textbf{0.90}&0.90\tabularnewline*
		~~2008Q1-2014Q4&\textbf{1.52}&1.62&\textbf{1.62}&1.64&\textbf{1.58}&1.61&\textbf{1.67}&1.81\tabularnewline*
		~~2015Q1-2019Q1&\textbf{0.80}&0.89&\textbf{0.85}&1.01&\textbf{0.82}&0.89&\textbf{0.85}&0.93\tabularnewline*
		\vspace{0.02 in}
		~\textcolor{black}{2005Q1-2019Q1}&\textcolor{black}{\textbf{1.00}}&\textcolor{black}{1.06}&\textcolor{black}{\textbf{1.05}}&\textcolor{black}{1.12}&\textcolor{black}{\textbf{1.02}}&\textcolor{black}{1.07}&\textcolor{black}{\textbf{1.08}}&\textcolor{black}{1.18}\tabularnewline
		\hline
		\rule{0pt}{2.5ex} 
		{\bfseries Forecast-1Q}&&&&&&&&\tabularnewline*
		\rule{0pt}{2.5ex} 
		~2005Q1-2007Q4&\textbf{1.05}&1.05&1.00&\textbf{0.93}&1.05&\textbf{1.03}&1.02&\textbf{0.97}\tabularnewline*
		~~2008Q1-2014Q4&\textbf{2.46}&2.46&\textbf{2.40}&2.51&\textbf{2.51}&2.59&\textbf{2.44}&2.59\tabularnewline*
		~~2015Q1-2019Q1&\textbf{0.85}&0.87&\textbf{0.87}&0.90&\textbf{0.88}&0.94&\textbf{0.89}&0.95\tabularnewline*
		\vspace{0.02 in}
		~\textcolor{black}{2005Q1-2019Q1}&\textcolor{black}{\textbf{1.42}}&\textcolor{black}{1.43}&\textcolor{black}{\textbf{1.39}}&\textcolor{black}{1.43}&\textcolor{black}{\textbf{1.44}}&\textcolor{black}{1.49}&\textcolor{black}{\textbf{1.42}}&\textcolor{black}{1.50}\tabularnewline
		\hline
		\rule{0pt}{2.5ex} 
		{\bfseries Forecast-2Q}&&&&&&&&\tabularnewline*
		\rule{0pt}{2.5ex} 
		~2005Q1-2007Q4&1.20&\textbf{1.16}&1.22&\textbf{1.14}&\textbf{1.26}&1.30&1.28&\textbf{1.22}\tabularnewline*
		~~2008Q1-2014Q4&2.91&\textbf{2.77}&\textbf{2.66}&2.76&\textbf{2.83}&2.85&\textbf{2.72}&2.72\tabularnewline*
		~~2015Q1-2019Q1&\textbf{0.84}&0.84&\textbf{0.83}&0.85&\textbf{0.83}&0.88&\textbf{0.81}&0.85\tabularnewline*
		\vspace{0.02 in}
		~\textcolor{black}{2005Q1-2019Q1}&\textcolor{black}{1.63}&\textcolor{black}{\textbf{1.57}}&\textcolor{black}{\textbf{1.52}}&\textcolor{black}{1.56}&\textcolor{black}{\textbf{1.60}}&\textcolor{black}{1.63}&\textcolor{black}{\textbf{1.54}}&\textcolor{black}{1.55}\tabularnewline
		\hline
\end{longtable}}

\subsubsection{LASSO-PC: gains from transformation}
Seeking to inspect directly the gains of using the principal component transformation on the (relaxed) data, a few more comparisons were made. First, in Table \ref{gfcf:RLvsPP} the results of forecasting GFCF are presented, when the number of preselected variables were fixed to 10, 15, 20, and 30. Note that the results in Tables \ref{gfcf:main1} and \ref{gfcf:main2} are generated by models with cross-validated hyperparameters. Therefore, the estimated number of significant variables may differ greatly during different time periods and between different models. In this case, we found that restricting the hyperparameter selection problem to select only a fixed amount of (the same) variables is useful. 

In Table \ref{gfcf:RLvsPP} we examine the results of two models: LP, where the preselected variables are transformed into principal components, and RL, where no transformation is made, only the coefficients are re-estimated following the Relaxed LASSO definition. In both cases the same variables are selected as significant. The results provide evidence that LP in some cases can improve the forecasting accuracy when compared with ordinary methods.

 Indeed, while some accuracy gains are visible in most cases, with 5\% significance the DM test suggests that LP15 and LP30 are able to generate significantly better 1-quarter forecasts than their respective RL versions. However, the results are less conclusive for nowcasts and 2-quarter forecasts due to larger estimated p-values of the DM test.
 
 To sum up, the improvement can be visible even on a relatively sparse number of variables selected, but the results suggest that the gains from using the PCA transformation are larger when more variables are added. This result is expected, since with larger samples we are likely to include more tightly related variables, thus allowing for a clearer extraction of the common factors. Indeed,  this becomes apparent in Section \ref{sub:leading}, where the 30 selected variables seem to form small correlated groups. While such groups may help extracting clearer signals through the PCA, the consequential collinearities can explain why in the RL case increasing the number of included variables did not improve the forecasting results.

In general, the RL associated results are based on two hyperparameters: $\lambda$ for the selection of variables, and $\phi$ for the amount of shrinkage applied. Seeking better understanding how the values of hyperparameters affect the forecasting performance we chose a set of indicators, preselected as optimal by the LASSO, and ran the pseudo-real-time forecasting exercise over the period of 2011Q1-2014Q4 for two cases: the first, where the rotation to the principal components is used and the second, where no transformation is applied and which roughly follows the Relaxed LASSO ideas. Finally, the case of the AdL was included, which uses a different set of variables for the prediction. 
The results presented in Figure \ref{gfcf:addGraph} show a slight improvement in both the average forecast accuracy -- relative mean RMSE comparing with benchmark model -- and a smaller standard deviation for many different values of the hyperparameter $\lambda$. These results provide further evidence that the use of principal components transformation might provide additional gains in forecast accuracy. 
\begin{figure}
	\includegraphics[width = 0.95\textwidth]{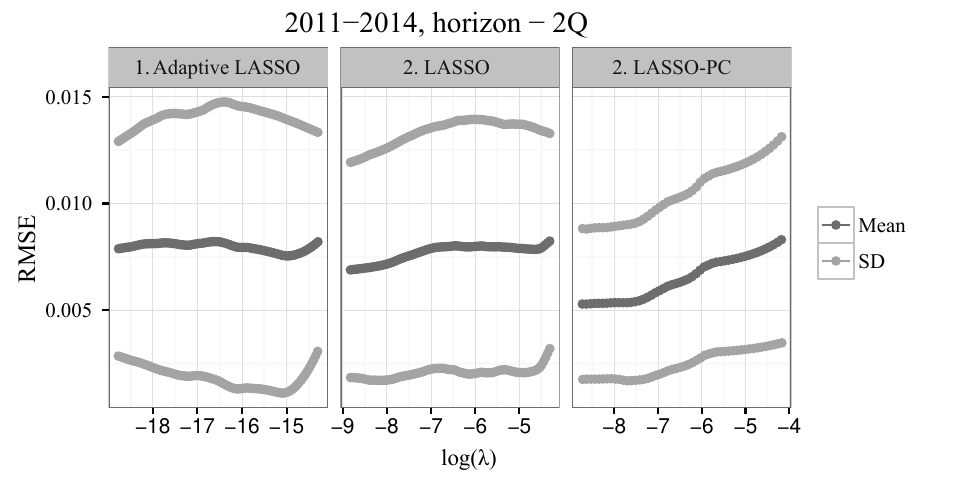}
	\caption{The results of forecasting accuracy during the pseudo-real-time experiments over 2011Q1-2014Q4 with the set of preselected variables being fixed for the whole period, and the numbers (1.) and (2.) enumerating the different sets of variables used. \label{gfcf:addGraph}}  
\end{figure}

\subsubsection{Sparse structures: implications from variable selection}
\label{sub:leading}
The main advantage of sparse structures is higher interpretability of the selected subsets of indicators. Figure \ref{gfcf:story1} depicts the top indicators frequently selected by the Relaxed LASSO during the pseudo-real-time experiments.

 We can see that there are several variables selected consistently during every period, which we interpret as the main drivers for explaining the investments in the US. Noteworthy, some of the variables form particular clusters, where one part describes the pre-crisis period while the other part gets significant after the crisis, indicating a possible structural break in the underlying information.

Among the most frequently selected variables are the data reflecting the situation in the market of dwellings: the number of employees and employment rate in construction, the number of building permits and building completions, and industrial production of construction supplies. These variables likely capture the Residential Investments part of the GFCF, following the evidence found in \cite{LUNSFORD2015276} and similar pre-screened variables detected for Italy in \cite{RollingWindow}.  

\setlongtables{\scriptsize
	\begin{longtable}{lccc}\caption{RMSE of models forecasts during pseudo-real experiments for GFCF. The bolded value is the smallest one for every row, 
			and for every block the last line denotes the total error of the period 2005-2019.}\label{gfcf:targeted} \tabularnewline\hline
		\multicolumn{1}{l}{\bfseries }&\multicolumn{1}{c}{\bfseries DirectPC}&\multicolumn{1}{c}{\bfseries \rule{0pt}{2.5ex}  \vspace{0.02 in} }&\multicolumn{1}{c}{\bfseries AggregatedPC}\tabularnewline*
		\hline
		\endfirsthead
		\caption{(Continued) RMSE of models forecasts during pseudo-real experiments for GFCF. The bolded value is the smallest one for every row, 
			and for every block the last line denotes the total error of the period 2005-2019.}
		\endhead
		\hline
		\endfoot
		\rule{0pt}{2.5ex}
		{\bfseries Backcast}&&&\tabularnewline*
		\rule{0pt}{2.5ex}
		~2005Q1-2007Q4&\textbf{0.53}&&0.53\tabularnewline*
		~~2008Q1-2014Q4&\textbf{0.47}&&0.47\tabularnewline*
		~~2015Q1-2019Q1&\textbf{0.57}&&0.57\tabularnewline*
		\vspace{0.02 in}
		~\textcolor{black}{2005Q1-2019Q1}&\textcolor{black}{\textbf{0.54}}&&\textcolor{black}{0.54}\tabularnewline
		\hline
		\rule{0pt}{2.5ex}
		{\bfseries Nowcast}&&&\tabularnewline*
		\rule{0pt}{2.5ex}
		~2005Q1-2007Q4&0.95&&\textbf{0.93}\tabularnewline*
		~~2008Q1-2014Q4&1.45&&\textbf{1.44}\tabularnewline*
		~~2015Q1-2019Q1&0.89&&\textbf{0.88}\tabularnewline*
		\vspace{0.02 in}
		~\textcolor{black}{2005Q1-2019Q1}&\textcolor{black}{1.03}&&\textcolor{black}{\textbf{1.01}}\tabularnewline

		\hline
		\rule{0pt}{2.5ex}
		{\bfseries Forecast-1Q}&&&\tabularnewline*
		\rule{0pt}{2.5ex}
		~2005Q1-2007Q4&1.21&&\textbf{1.16}\tabularnewline*
		~~2008Q1-2014Q4&2.55&&\textbf{2.22}\tabularnewline*
		~~2015Q1-2019Q1&0.95&&\textbf{0.91}\tabularnewline*
		\vspace{0.02 in}
		~\textcolor{black}{2005Q1-2019Q1}&\textcolor{black}{1.51}&&\textcolor{black}{\textbf{1.36}}\tabularnewline
		\hline
		\rule{0pt}{2.5ex}
		{\bfseries Forecast-2Q}&&&\tabularnewline*
		\rule{0pt}{2.5ex}
		~2005Q1-2007Q4&1.39&&\textbf{1.29}\tabularnewline*
		~~2008Q1-2014Q4&\textbf{2.76}&&2.76\tabularnewline*
		~~2015Q1-2019Q1&0.89&&\textbf{0.83}\tabularnewline*
		\vspace{0.02 in}
		~\textcolor{black}{2005Q1-2019Q1}&\textcolor{black}{1.61}&&\textcolor{black}{\textbf{1.57}}\tabularnewline
		\hline
\end{longtable}}

Second, the productive investment decisions are often determined by the outlook in the related labour market, capturing the business cycle dynamics over different regions, industries and age groups. Such a broad scope may help refining the signals from the labour market. On top of that, Industrial Production Indices proxy the current state of the economic cycle and the expectations in aggregated demand. 

   In this regard, the use of principal components to extract the underlying common factor explaining the regional diversity in economic activity is deemed necessary for more efficient estimation of the model parameters. These results, however, contrast with \cite{Bok2018}, who argue that disaggregated variables, such as employment by age or industry, show no substantial gains in prediction accuracy.

\begin{figure}
	\includegraphics[width = 1\textwidth]{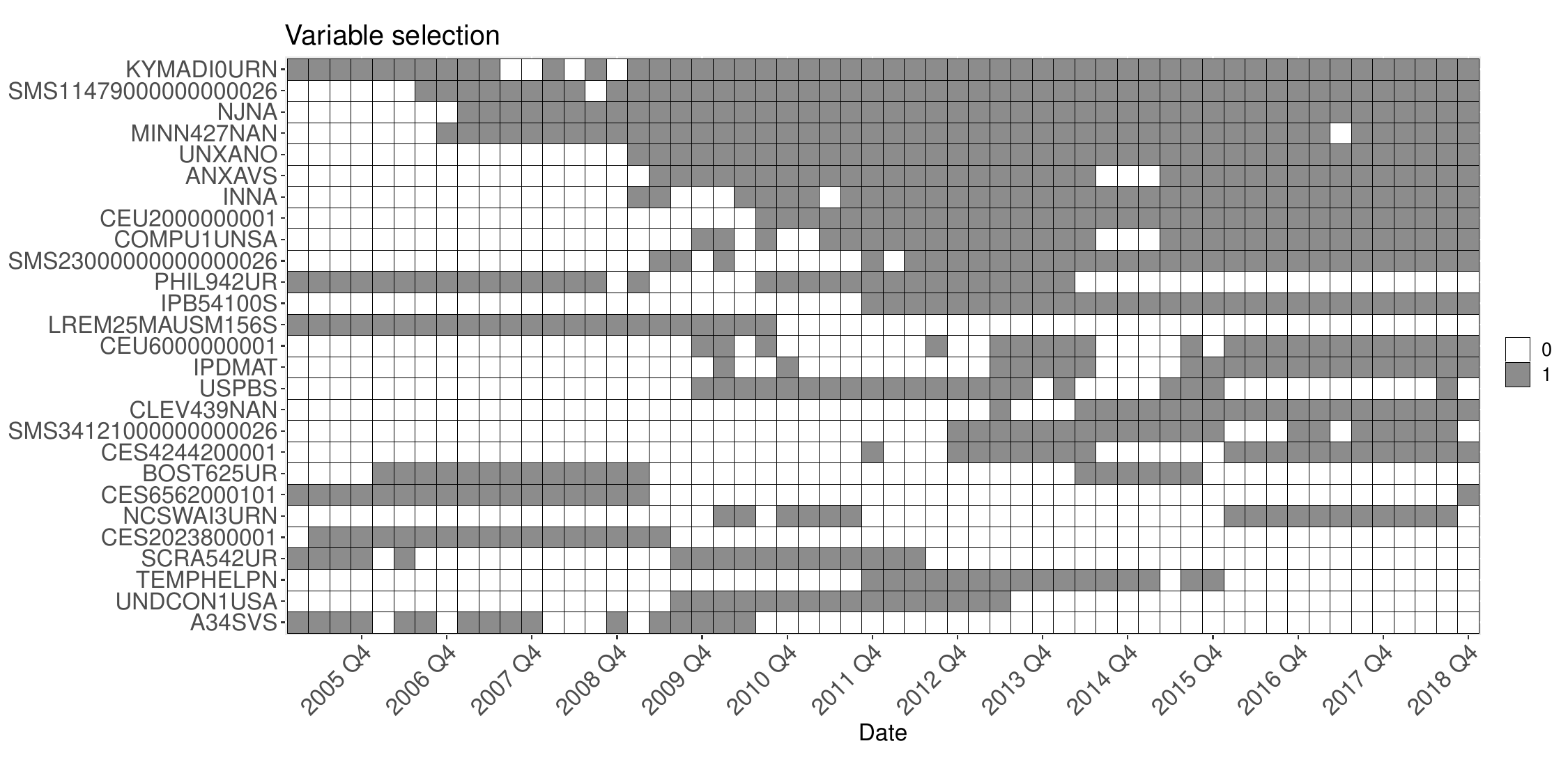}
	
	\caption{Most often selected variables during the expanding window pseudo-real-time experiments for GFCF over 2005Q1-2018Q4. The number of times selected denotes only the number of the same variables selected (e.g., the variable and a one-quarter lag) but not which lag was most often selected. Acronyms described in Appendix, Table \ref{variables:expanding}.  \label{gfcf:story1}}
\end{figure}

\subsubsection{Forecast aggregation}
To evaluate the ideas of forecast aggregation, briefly discussed in Section \ref{PCA}, under real data, we conduct the following additional experiment. In this experiment, first, we preselect a set of significant variables by the LASSO during every quarter of the pseudo-real-time exercise. Second, we consider the post-LASSO model, which applies an OLS regression, using the first five principal components ordered by their variance explained\footnote{It is often found in the literature (e.g., \cite{doi:10.1111/1468-0262.00273}) that a small number of principal components is usually enough when the initial data sample is relatively sparse. In our case, we noticed that typically around 10 to 30 variables are selected during the exercise, so this number seems optimal since it is not too large for efficient OLS estimation and not too small to omit underlying data. Also, since the variables are preselected by the LASSO, it is likely that principal components, explaining the most variance, will be the most significant in the OLS estimation.} and treating the assessed factors as explanatory variables. However, as discussed in Section \ref{PCA}, we examine two predicting approaches: the first -- by fitting an appropriate ARMA model for each of the components and forecasting them directly ({DirectPC}); and the second -- by forecasting the preselected variables and aggregating their forecasts ({AggregatedPC}). Table \ref{gfcf:targeted} presents the resulting performance of both of these methods, and a few conclusions arise. 

First, we can see that the nowcasting performance is as expected very similar, with the aggregated method being able to explain the crisis period more accurately than the direct approach. Indeed, since some monthly information is already known during the nowcasted quarter and the factors compared are the same, such comparison essentially depends on the method used to fill the ragged edges\footnote{In this exercise the ragged edges were filled using the Holt-Winters approach. It is not the most commonly used method for such a problem, but we have found it producing adequate results. ARMA models are also a good alternative, but we did not want to have coinciding nowcasts with the ones from the {AggregatedPC}. }.

 Second, the forecasting accuracy for most of the periods visibly improved using the aggregated forecast method, with the highest differences during the crisis period of 2008Q1-2014Q4. An inferior forecasting performance by direct approach can likely be caused by underestimating the complexity of the extracted factors. Even if the amount of sample data is relatively large, it may be not enough for efficient estimation of all ARMA parameters. Although the same holds for forecasting the preselected variables, the aggregation of their forecasts appears to improve the results. In this case, the most significant gains were seen for the 1-quarter ahead forecasts. Besides, following the discussion in Section \ref{PCA}, we expect more persistent dynamics with a broader scope of information explaining each of the subcomponents.
 
  The benefits of the forecast aggregation are twofold: the creation of a more complex dynamics of the final forecasts than by forecasting directly; and a self-correction by smoothing out inaccurate individual forecasts.

A few additional observations follow from these results. First, because of the complexity of modelling each component, the aggregation is feasible for only a small subset of variables and is not applicable for large and dense problems. However, the complexity of the problem with preselected (targeted) predictors essentially reduces to the complexity of the Relaxed LASSO method. Second, comparing the results from Table \ref{gfcf:main1} with Table \ref{gfcf:targeted}, it can be seen that since the variables preselected in both cases are the same, the Post-LASSO solution with using only the first few extracted principal components can even lead to comparable results with original LASSO and LP.

\subsection{Rolling window: US Gross Fixed Capital Formation}
In this section, we repeat the exercise by switching to a 13-year rolling window. The size of the window reflects the likely occurrence of structural breaks within both standard business cycle frequencies up to 8 years and supply-side driven medium run frequencies up to 13 years. The primary goal of the rolling window is to inspect the impact of old historical information, e.g., continuously chosen by the LASSO as significant only because they help to explain the historical data at the start of the sample but are less useful for the most recent predictions.

\subsubsection{Main findings}
The main results are presented in Tables \ref{gfcf:rolling2} and \ref{gfcf:rolling22}. As in the expanding forecast window, we can see that most analyzed models can outperform the dense benchmarks. The BNDF model generates the most accurate nowcasts, while L0L1 generates respectively the most accurate 1- and 2-quarter forecasts. 

Most variants of the LASSO generates significantly more accurate nowcasts as suggested by the GW test (see Table \ref{gfcf:rolling22}) when compared with the ARMA and DF, SW benchmarks, however, the evidence against BNDF is lacking. On the other hand, the evidence is more visible for the 1- and 2-quarter ahead forecasts. 

Overall results of the nowcasting performance tightly depend on the variable selection or the accuracy of signal extraction. Therefore, the differences are more visible when comparing dense methods with sparse, bearing in mind that BNDF uses the same variable selection to LP, LPX and RL methods. On the other hand, comparing LP with BNDF directly we are inspecting the gains from the additional shrinkage and factor selection, brought by LP. Finally, examining the forecasting performance, the different approaches for variable forecasting comes in play. For longer, 2-quarter ahead forecasts, we observe the higher gains of ARMA vs Kalman filter approaches, while the 1-quarter forecasts should also capture the forecast aggregation gains, following the discussion around Table \ref{gfcf:targeted} and the ideas from Section \ref{PCA}.

Noteworthy, when comparing the main results with the ones from the expanding window (see Tables \ref{gfcf:main1} and \ref{gfcf:rolling2}), LP overall accuracy for the full data sample is very similar in both cases. However, by inspecting the nowcasting performance during different spliced periods, it appears that the rolling window setup leads to lower accuracy during the crisis period, but higher onward from 2015Q1. This finding provides evidence for the possible effect of structural changes in the composition of the selected data. In essence, the loss of accuracy during the crisis period suggests that the chosen 13-year window may be too small since the models miss information during the first periods of the estimation. To test the idea, we have repeated the estimation with a 16-year rolling window and were able to increase the nowcasting performance. Although beyond the scope of the exercise, this finding reveals that the size of the rolling window may be treated as a hyperparameter for structural change flexibility and can be further explored, in combination with the resulting sparse structures.

\subsubsection{Sparse structures}
In Figure \ref{gfcf:roll:vars}, the list of top variables, preselected by the LASSO during the forecasting exercise, is presented. It can be seen that similar to the results from the expanding window estimation, most of the consistently selected indicators explain the construction and housing sectors in the US. The employment rate in the construction sector, together with numbers on building permissions and building completions, complemented by the Consumer Price Index in the housing sector provide a rather detailed view on the situation in the housing market. 

By comparing the differences between rolling and expanding windows, two main effects occur. The first effect is the inclusion of more informative variables that become accessible at later periods, e.g., additional indicators for the construction sector. The second is the flexible adjustment to structural changes, e.g., the shift of regions for the employment variables that can be related to different active economic activities, or the exclusion of information that was relevant only during the crisis period.

 \setlongtables{\scriptsize
 	\setlength\tabcolsep{4.5pt}
 	\begin{longtable}{lcccc|cccc|cccc|ccccccccccccccc}\caption{RMSE of models forecasts during pseudo-real experiments with rolling window for US GFCF.  The labels B, N, Q1 and Q2 denotes the forecast horizons -- Backcast, Nowcast, Forecast-1Q and Forecast-2Q, respectively. The bolded value is the smallest one for every column, and every block corresponds to a splitted time period, while the last block shows the results for the full period ove 2005-2019. \label{gfcf:rolling2}} \tabularnewline \hline\hline
 		\bfseries \multirow{2}{*}{Model}&\multicolumn{4}{c}{\bfseries 2005Q1-2007Q4}&\multicolumn{4}{c}{\bfseries 2008Q1-2014Q4}&\multicolumn{4}{c}{\bfseries 2015Q1-2019Q1}&\multicolumn{4}{c}{\bfseries 2005Q1-2019Q1 \rule{0pt}{2.5ex} } \tabularnewline*
 		&\bfseries B&\bfseries N&\bfseries Q1&\bfseries Q2&\bfseries B&\bfseries N&\bfseries Q1&\bfseries Q2&\bfseries B&\bfseries N&\bfseries Q1&\bfseries Q2&\bfseries B&\bfseries N&\bfseries Q1&\bfseries Q2 \rule{0pt}{2.5ex} \tabularnewline*
 		\hline
 		\endfirsthead
 		\caption{(Continued) RMSE of models forecasts during pseudo-real experiments with rolling window for US GFCF.  The labels B, N, Q1 and Q2 denotes the forecast horizons -- Backcast, Nowcast, Forecast-1Q and Forecast-2Q, respectively. The bolded value is the smallest one for every column, and every block corresponds to a splitted time period, while the last block shows the results for the full period ove 2005-2019. \label{gfcf:rolling2}}
 		\endhead
 		\hline 
 		\endfoot
 		L \rule{0pt}{2.5ex} &0.49&0.86&1.02&1.12&0.85&1.84&2.62&2.95&0.44&0.77&0.90&0.93&\textcolor{black}{0.57}&\textcolor{black}{1.10}&\textcolor{black}{1.50}&\textcolor{black}{1.67} \\
 		LP&0.31&0.80&0.98&1.15&0.47&1.57&2.56&3.14&0.33&0.73&\textbf{0.89}&0.94&\textcolor{black}{0.35}&\textcolor{black}{0.98}&\textcolor{black}{1.46}&\textcolor{black}{1.76} \\ 
 		LPX&0.28&0.83&1.02&1.14&0.38&1.73&2.68&3.21&0.32&0.71&0.91&\textbf{0.92}&\textcolor{black}{0.32}&\textcolor{black}{1.04}&\textcolor{black}{1.52}&\textcolor{black}{1.78} \\ 
 		RL&0.19&0.77&0.97&1.14&0.38&1.81&2.66&3.16&0.29&0.79&0.92&0.92&\textcolor{black}{0.28}&\textcolor{black}{1.09}&\textcolor{black}{1.52}&\textcolor{black}{1.76} \\ 
 		SqL&0.55&0.84&1.04&1.16&0.69&1.78&2.67&2.96&0.44&0.77&0.90&0.92&\textcolor{black}{0.52}&\textcolor{black}{1.08}&\textcolor{black}{1.51}&\textcolor{black}{1.68} \\ 
 		SqP&0.38&0.84&1.00&1.16&0.43&1.74&2.62&2.99&0.36&0.75&0.91&0.94&\textcolor{black}{0.37}&\textcolor{black}{1.06}&\textcolor{black}{1.49}&\textcolor{black}{1.69} \\ 
 		AdP&0.07&0.71&0.82&1.04&0.15&1.87&2.54&2.94&0.09&0.74&0.94&0.96&\textcolor{black}{0.10}&\textcolor{black}{1.09}&\textcolor{black}{1.45}&\textcolor{black}{1.66} \\ 
 		AdL&0.10&0.72&0.84&1.03&0.19&1.91&2.58&2.91&0.12&0.75&0.94&0.93&\textcolor{black}{0.14}&\textcolor{black}{1.12}&\textcolor{black}{1.47}&\textcolor{black}{1.64} \\ 
 		AdRL&\textbf{0.04}&\textbf{0.63}&\textbf{0.81}&\textbf{1.02}&\textbf{0.07}&1.79&2.53&2.90&\textbf{0.06}&0.75&0.96&0.96&\textcolor{black}{\textbf{0.06}}&\textcolor{black}{1.05}&\textcolor{black}{1.45}&\textcolor{black}{1.65} \\ 
 		L0L1&0.77&1.17&1.25&1.23&1.01&1.76&\textbf{2.12}&\textbf{2.51}&0.58&0.86&0.91&0.94&\textcolor{black}{0.74}&\textcolor{black}{1.19}&\textcolor{black}{\textbf{1.35}}&\textcolor{black}{\textbf{1.52}} \\ 
 		L0L1p&0.75&1.15&1.27&1.23&0.98&1.83&2.15&2.57&0.51&0.91&0.93&0.93&\textcolor{black}{0.70}&\textcolor{black}{1.22}&\textcolor{black}{1.37}&\textcolor{black}{1.54} \\ 
 		L0&0.65&0.99&1.18&1.26&0.93&1.82&2.26&2.69&0.82&0.94&0.99&1.02&\textcolor{black}{0.81}&\textcolor{black}{1.21}&\textcolor{black}{1.43}&\textcolor{black}{1.61} \\ 
 		L0p&0.65&0.98&1.17&1.26&0.90&1.82&2.26&2.69&0.82&0.93&0.99&1.02&\textcolor{black}{0.80}&\textcolor{black}{1.21}&\textcolor{black}{1.43}&\textcolor{black}{1.61} \\ 
 		L0L2&0.75&0.99&1.26&1.23&1.01&1.94&2.50&2.61&0.56&0.81&0.94&0.94&\textcolor{black}{0.73}&\textcolor{black}{1.20}&\textcolor{black}{1.51}&\textcolor{black}{1.55} \\ 
 		L0L2p&0.75&1.00&1.26&1.24&0.94&1.99&2.61&2.70&0.54&0.84&0.93&0.92&\textcolor{black}{0.70}&\textcolor{black}{1.23}&\textcolor{black}{1.54}&\textcolor{black}{1.58} \\ 
 		\hline
 		ARMA \rule{0pt}{2.5ex} &0.87&0.90&1.03&1.17&2.19&2.50&3.45&3.60&0.99&1.08&1.13&1.29&\textcolor{black}{1.34}&\textcolor{black}{1.49}&\textcolor{black}{1.93}&\textcolor{black}{2.01} \\ 
 		BNDF&0.54&0.83&1.06&1.20&0.75&\textbf{1.48}&2.72&3.65&0.55&\textbf{0.68}&0.89&0.97&\textcolor{black}{0.56}&\textcolor{black}{\textbf{0.94}}&\textcolor{black}{1.55}&\textcolor{black}{1.99} \\ 
 		DF&1.23&1.33&1.43&1.52&1.02&2.35&3.18&3.23&0.77&1.01&1.10&1.09&\textcolor{black}{0.89}&\textcolor{black}{1.47}&\textcolor{black}{1.84}&\textcolor{black}{1.88} \\ 
 		SW&1.31&1.42&1.53&1.59&1.74&2.32&3.40&4.15&0.86&0.95&1.01&1.14&\textcolor{black}{1.18}&\textcolor{black}{1.46}&\textcolor{black}{1.92}&\textcolor{black}{2.28} \\

 \end{longtable}}

\setlongtables{\scriptsize
	\begin{longtable}{lclllclllclllclll}\caption{Relative RMSE to a set of benchmark models, during pseudo-real experiments for GFCF over the full period of 2005-2019. The bolded value corresponds to the best performance per forecast horizon. Asterix indicates significant differences in forecasting accuracy, when compared with the according benchmark, based on GW test with 5\% significance.  \label{gfcf:rolling22}} \tabularnewline*
		\hline\hline
		\multicolumn{1}{c}{\bfseries Model}&\multicolumn{1}{c}{\bfseries \rule{0pt}{2.5ex}  \vspace{0.02 in} }&\multicolumn{3}{c}{\bfseries ARMA}&\multicolumn{1}{c}{\bfseries }&\multicolumn{3}{c}{\bfseries BNDF}&\multicolumn{1}{c}{\bfseries }&\multicolumn{3}{c}{\bfseries DF}&\multicolumn{1}{c}{\bfseries }&\multicolumn{3}{c}{\bfseries SW}\tabularnewline
		\cline{3-5} \cline{7-9} \cline{11-13} \cline{15-17}
		\multicolumn{1}{c}{\rule{0pt}{2.5ex}  \vspace{0.02 in}}&\multicolumn{1}{c}{}&\multicolumn{1}{c}{\textbf{N}}&\multicolumn{1}{c}{\textbf{Q1}}&\multicolumn{1}{c}{\textbf{Q2}}&\multicolumn{1}{c}{}&\multicolumn{1}{c}{\textbf{N}}&\multicolumn{1}{c}{\textbf{Q1}}&\multicolumn{1}{c}{\textbf{Q2}}&\multicolumn{1}{c}{}&\multicolumn{1}{c}{\textbf{N}}&\multicolumn{1}{c}{\textbf{Q1}}&\multicolumn{1}{c}{\textbf{Q2}}&\multicolumn{1}{c}{}&\multicolumn{1}{c}{\textbf{N}}&\multicolumn{1}{c}{\textbf{Q1}}&\multicolumn{1}{c}{\textbf{Q2}}\tabularnewline
		\hline
		\endfirsthead\caption[]{\em (continued)} \tabularnewline
		\hline
		\multicolumn{1}{c}{\bfseries Model}&\multicolumn{1}{c}{\bfseries \rule{0pt}{2.5ex}  \vspace{0.02 in}}&\multicolumn{3}{c}{\bfseries ARMA}&\multicolumn{1}{c}{\bfseries }&\multicolumn{3}{c}{\bfseries BNDF}&\multicolumn{1}{c}{\bfseries }&\multicolumn{3}{c}{\bfseries DF}&\multicolumn{1}{c}{\bfseries }&\multicolumn{3}{c}{\bfseries SW}\tabularnewline
		\cline{3-5} \cline{7-9} \cline{11-13} \cline{15-17}
		\multicolumn{1}{c}{\rule{0pt}{2.5ex}  \vspace{0.02 in}}&\multicolumn{1}{c}{}&\multicolumn{1}{c}{\textbf{N}}&\multicolumn{1}{c}{\textbf{Q1}}&\multicolumn{1}{c}{\textbf{Q2}}&\multicolumn{1}{c}{}&\multicolumn{1}{c}{\textbf{N}}&\multicolumn{1}{c}{\textbf{Q1}}&\multicolumn{1}{c}{\textbf{Q2}}&\multicolumn{1}{c}{}&\multicolumn{1}{c}{\textbf{N}}&\multicolumn{1}{c}{\textbf{Q1}}&\multicolumn{1}{c}{\textbf{Q2}}&\multicolumn{1}{c}{}&\multicolumn{1}{c}{\textbf{N}}&\multicolumn{1}{c}{\textbf{Q1}}&\multicolumn{1}{c}{\textbf{Q2}}\tabularnewline
		\hline
		\endhead
		\hline
		\endfoot
		
		AdL&\rule{0pt}{2.5ex} &0.75*&0.76*&0.82*&&1.20*&0.95&0.83&&1.03&0.95*&0.84*&&0.77*&0.77*&0.72*\tabularnewline
		AdP&&0.73*&0.75*&0.83*&&1.17&0.94&0.84&&1.00&0.93*&0.85*&&0.75*&0.76*&0.73*\tabularnewline
		AdRL&&0.71*&0.75*&0.82*&&1.13&0.94&0.83&&0.97&0.94*&0.85*&&0.72*&0.76*&0.72*\tabularnewline
		L&&0.74*&0.78*&0.83*&&1.18&0.97&0.84&&1.01&0.97&0.86*&&0.76*&0.78*&0.73*\tabularnewline
		L0&&0.81&0.74*&0.80*&&1.29*&0.93&0.81&&1.11&0.92&0.83&&0.83&0.75&0.71*\tabularnewline
		L0L1&&0.80&\textbf{0.70*}&\textbf{0.75*}&&1.27*&\textbf{0.88}&\textbf{0.76}&&1.09&\textbf{0.87}&\textbf{0.78*}&&0.82&\textbf{0.71*}&\textbf{0.67*}\tabularnewline
		L0L1p&&0.82&0.71*&0.77*&&1.31*&0.89&0.77&&1.12&0.88&0.79&&0.84&0.71*&0.68*\tabularnewline
		L0L2&&0.80&0.78*&0.77*&&1.28*&0.98&0.78&&1.10&0.97&0.80*&&0.82*&0.79*&0.68*\tabularnewline
		L0L2p&&0.82&0.80*&0.79*&&1.31*&1.00&0.79&&1.13&1.00&0.81&&0.84&0.80*&0.69*\tabularnewline
		L0p&&0.81&0.74*&0.80*&&1.29*&0.92&0.81&&1.11&0.92&0.83&&0.83&0.74&0.71*\tabularnewline
		LP&&\textbf{0.66*}&0.76*&0.87*&&\textbf{1.05}&0.95&0.88&&\textbf{0.90}&0.94*&0.90*&&\textbf{0.67*}&0.76*&0.77*\tabularnewline
		LPX&&0.70*&0.79*&0.88*&&1.11&0.99&0.89&&0.95&0.98&0.91*&&0.71*&0.79*&0.78*\tabularnewline
		RL&&0.73*&0.79*&0.88*&&1.17*&0.98&0.89&&1.00&0.98&0.90*&&0.75*&0.79*&0.77*\tabularnewline
		SqL&&0.72*&0.79*&0.83*&&1.15*&0.98&0.84&&0.99&0.98&0.86*&&0.74*&0.79*&0.74*\tabularnewline
		SqP&&0.71*&0.78*&0.84*&&1.14*&0.97&0.85&&0.97&0.96&0.87*&&0.73*&0.78*&0.74*\tabularnewline
		\hline
\end{longtable}}

\noindent%
\begin{minipage}{\linewidth}
	\makebox[\linewidth]{
		\includegraphics[width = 0.99\textwidth]{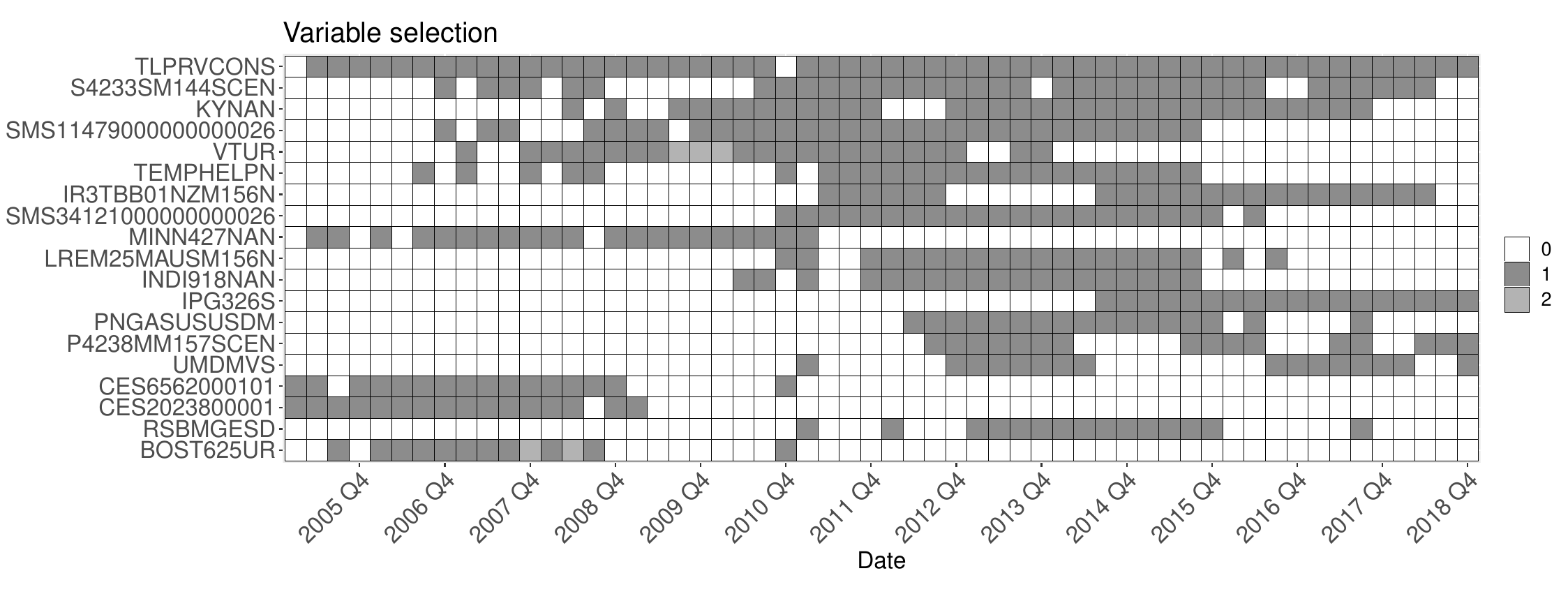}
	}
	\captionof{figure}{Most often selected variables by the LASSO during the rolling window pseudo-real-time forecasting exercise for the GFCF. The number of times selected denotes only the number of the same variables selected (e.g., the variable and a one-quarter lag) but not which lag was most often selected. Acronyms described in Appendix, Table \ref{variables:rolling} }\label{gfcf:roll:vars}
\end{minipage}

\subsection{US Private Final Consumption Expenditure}

Compared with the investments, the behaviour of private consumption is quite different. First, it tends to show a much more stable growth than investments. Second, it does not immediately respond to the various stages of the business cycle -- the private consumption tends to take the momentum only when the expansion of the current cycle is well underway, with reaching the peak after the cycle. Therefore, it is easier to reflect various shocks in the economy when generating nowcasts for private consumption, since in particular markets some of the shocks could be perceived much earlier. Additionally, for nowcasting, it is especially convenient to use available ``hard'' monthly real personal consumption indicators released with a relatively small publication lag. Furthermore, this finding highlights the importance of accurate individual forecasting of such monthly indicators. It is very likely, that 1- and 2-quarter forecasts of private consumption would be greatly improved if the forecasts of ``hard'' monthly indicators would be generated by employing more sophisticated models, capable of including more explanatory information than benchmark models.

The results of forecasting the PFCE are presented in Tables \ref{table:final} and \ref{table:final2} of the Appendix \ref{appendix:tables} over a rolling 13-year window.  The most accurate nowcasts are produced by the L0L1P models, with rather similar performance across most LASSO variants. Besides, the most accurate 1- and 2-quarter forecasts are generated by the BNDF and LP, respectively.  We note that while the nowcasting performance is rather similar to the benchmark models (as suggested by GW test, see Table \ref{pfce:tables}), most LASSO variants can significantly outperform DF and SW benchmarks for nowcasts and 1-quarter forecasts. 

Figure \ref{cons:roll:vars} presents the top monthly variables most often preselected by the LASSO as significant. When inspecting the results, we find that the most frequently selected are the ``hard'' monthly indicators of real personal consumption expenditure (expenditures on goods and services) as was expected. Indeed, statistical agencies themself use ``hard'' data as the primary sources for their preliminary nowcasts, confirming that LASSO can identify the main leading indicators from a large set of available information.

\begin{figure}
	\includegraphics[width = 0.99\textwidth]{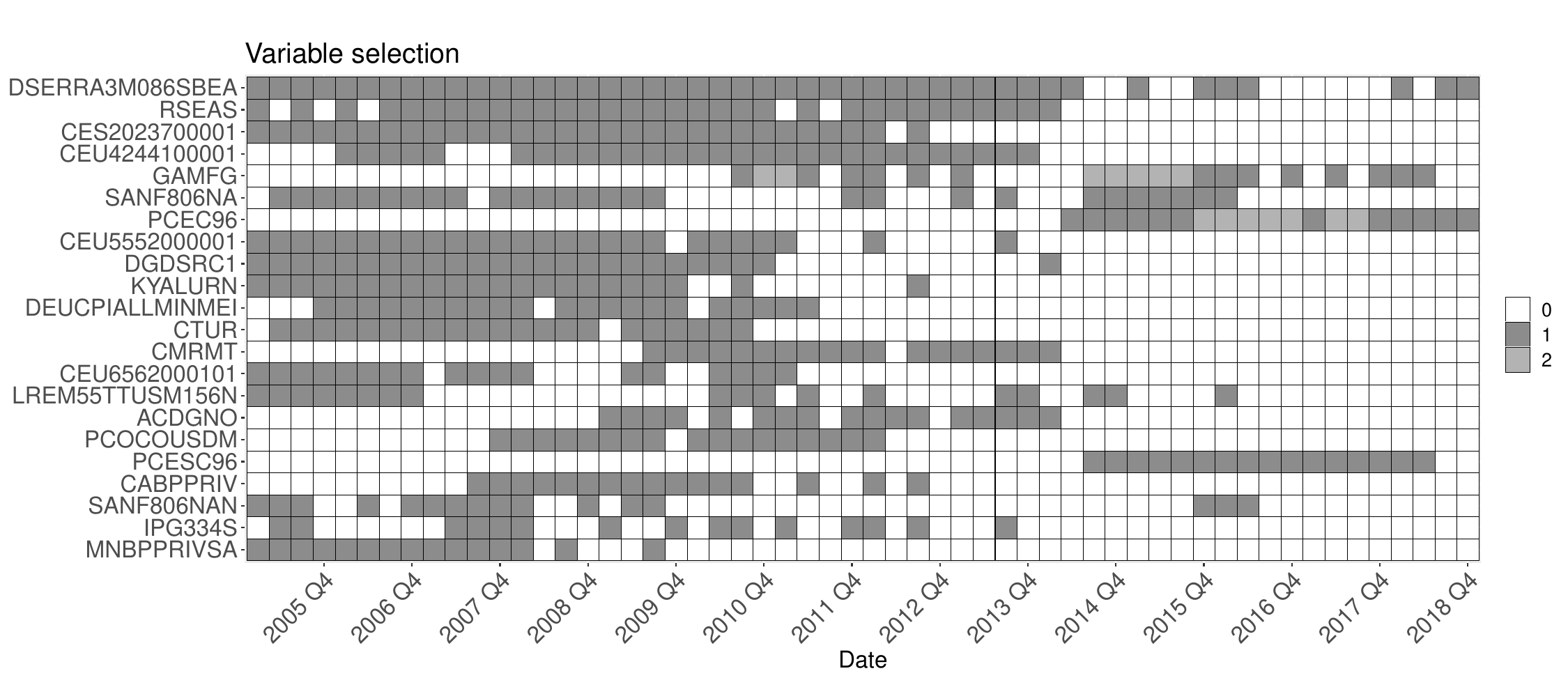}
	\caption{Most often selected variables by the LASSO during the rolling window pseudo-real-time forecasting exercise for the PFCE. The number of times selected denotes only the number of the same variables selected (e.g., the variable and a one-quarter lag) but not which lag was most often selected. Acronyms described in Appendix, Table \ref{variables:rolling:pfce} } \label{cons:roll:vars}
\end{figure}

\subsection{US International Trade}

Similarly to investments, the exports and imports of goods and services are more volatile than the aggregate GDP. The cyclical properties of international trade are quite appealing and follow from the balance of two forces. First, the 'loving-variety' economic agents that seek to smooth consumption using international trade. Second, the additional cyclical variability comes from the investments, that are permitted by the international capital flows. Although there usually exists a strong co-movement between exports and imports, the response to the shocks may have an opposite or supplementary impact on exports and imports. For instance, appreciation of the real effective exchange rate can be expected to decrease exports due to the reduced price competitiveness, but increase imports by lowering the relative import prices. Some strong demand shocks, on the contrary, might pass through the global economy leading to acceleration or are affected by global boom-bust cycles in economic activity. For example, during acceleration, an initial increase in imports due to the hike in the US domestic demand results in growing foreign exports and foreign income, which in turn increase domestic exports. As in the case with private consumption, the nowcasting of exports and imports is simplified by the existence of ``hard'' monthly indicators of external trade, published with a small delay. 

\begin{figure}

	\includegraphics[width = 0.99\textwidth]{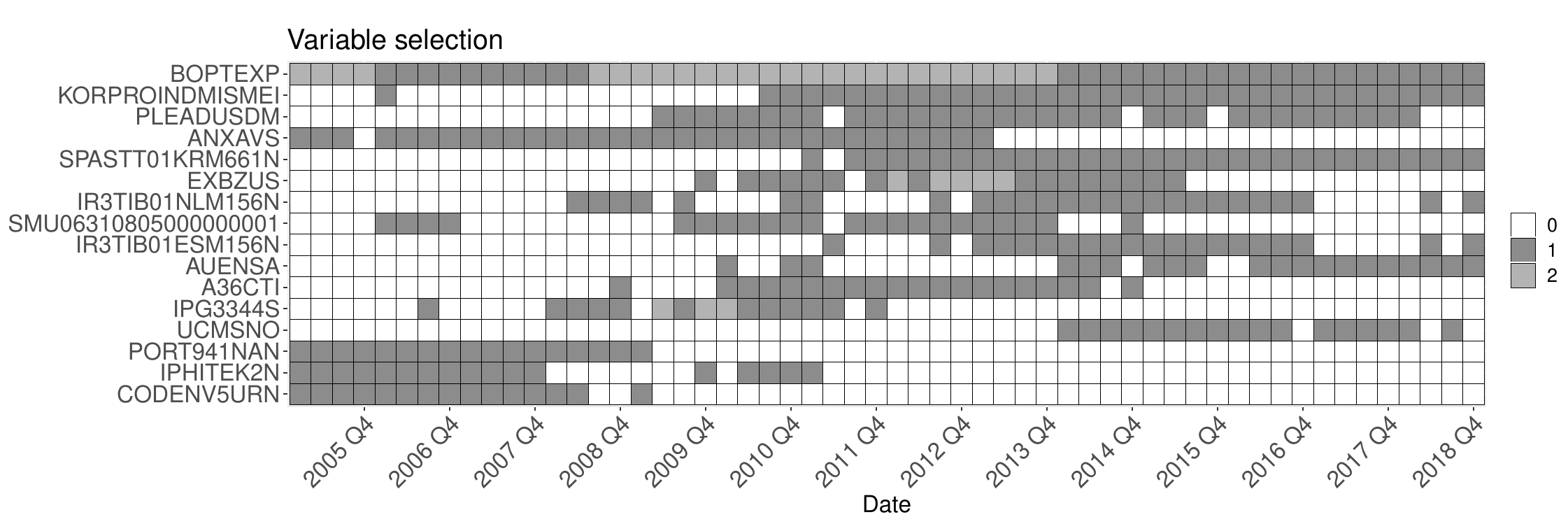}
	\caption{Most often selected variables by the LASSO during the rolling window pseudo-real-time forecasting exercise for the Exports. The number of times selected denotes only the number of the same variables selected (e.g., the variable and a one-quarter lag) but not which lag was most often selected. Acronyms described in Appendix, Table \ref{variables:rolling:exp} \label{exp:roll:vars}}  
\end{figure}

The results of forecasting Exports and Imports are presented in Table \ref{table:final} of the Appendix \ref{appendix:tables} over a rolling 13-year window. 

When nowcasting the Exports, all variants of the LASSO can outperform the ARMA models, however, are very close to the nowcasts of BNDF, lacking evidence of significance by the GW test. Despite this, the lowest RMSE are generated by the suggested in this paper LPX method. The latter provides further evidence that additional forecasting accuracy can follow from using a mixture between the two methods, where both the principal components and the original preselected data are included in the model. Finally, this finding shows that tailoring the loading matrix can help to reduce the noise from the data, as discussed in Section \ref{sparsePCA}.

 When forecasting 1- and 2-quarters ahead, the results are even closer between all of the models, with BNDF having the most accuracy by a small margin. However, these results are expected due to the existence of ``hard'' monthly indicators -- the nowcasting results highly depends on the accurate variable selection part, while the forecasts reflect the model structure and the accuracy of the monthly indicator forecasts. Since the forecasts are likely lead by the monthly Exports of Goods and Services variable (see Figure \ref{exp:roll:vars}, Table \ref{variables:rolling:exp}), we do not expect significant forecast accuracy improvements due to forecast aggregation as per the discussion in Section \ref{lassopcamotivation}, hence the very similar results between all of the model forecasts.

When nowcasting the Imports, AdRL shows the best result, significantly outperforming the DF model, as suggested by the GW test. Additionally, SqL generates the best 1-quarter forecasts, while SqP generates the best 2-quarter forecasts. The GW test results are weaker for the forecasts, suggesting that the large observed RMSE differences may be caused by a few distinct points, most likely during the crisis period.

\begin{figure}
		\includegraphics[width = 0.99\textwidth]{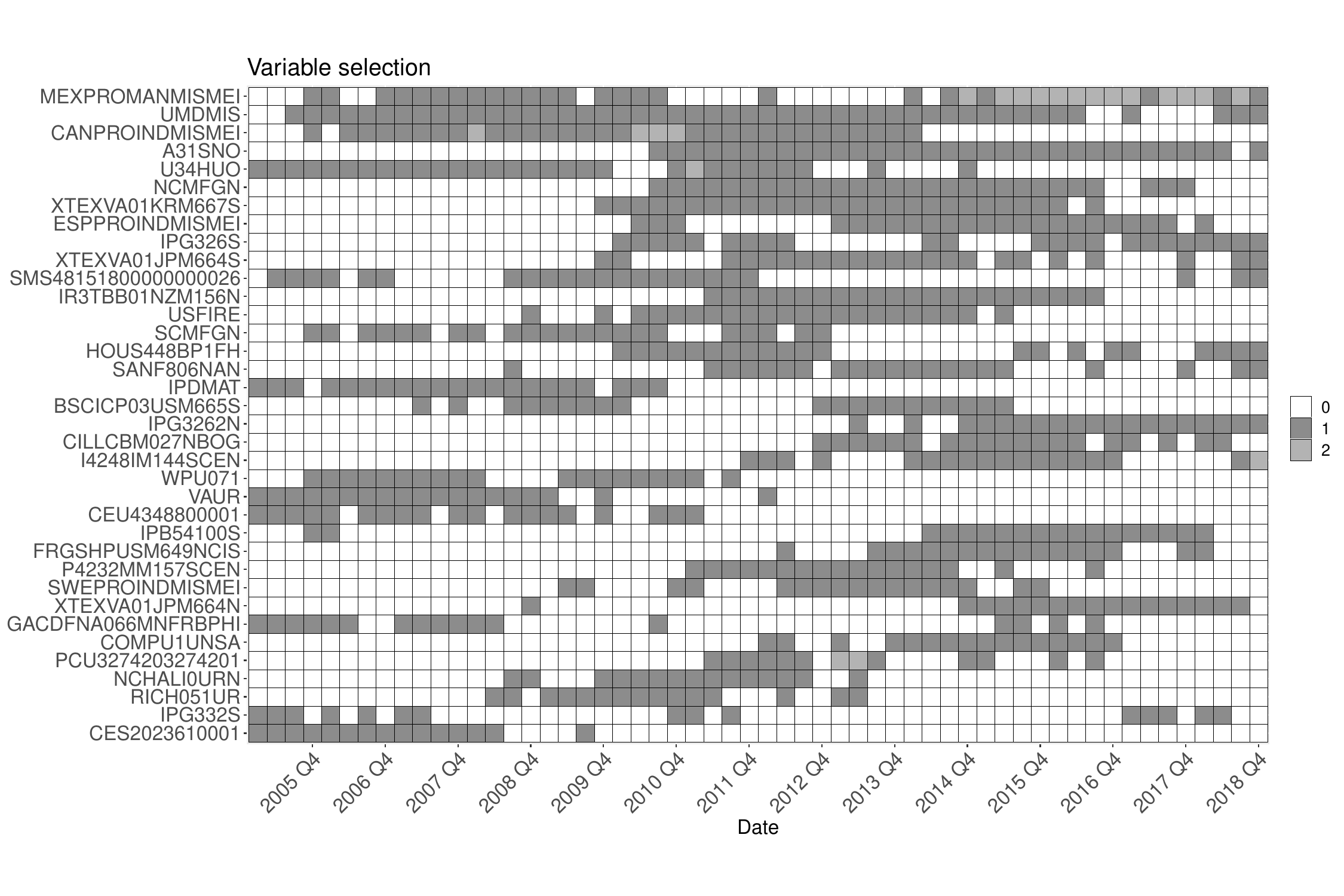}
	\caption{Most often selected variables by the LASSO during the rolling window pseudo-real-time forecasting exercise for the Imports. The number of times selected denotes only the number of the same variables selected (e.g., the variable and a one-quarter lag) but not which lag was most often selected. Acronyms described in Appendix, Table \ref{variables:rolling:imp} \label{imp:roll:vars}}     
\end{figure}

\subsection{European countries}

As a robustness check, we repeat the estimation procedure for four largest Euro Area economies that are also the main contributors to the EU budget: Germany, France, Italy and Spain. Replicating the estimation, we cover the same four main components of the GDP accounted by the expenditure approach. Although, due to the varying availability of the historical data, we had to make certain adjustments. As with the US, it is of interest to cover the financial crisis period in the experiments. However, for some series, the available GDP data starts later than 1992Q1 -- in such cases, the rolling window is set to expand until it reaches the previously defined 13-year window. 

The results are presented in Appendix \ref{appendix:tables:eu}. As can be expected, no single forecasting approach dominates in the ``horse race''. However, we note that unlike with the results for the US data, there were cases when the dense factor methods were able to outperform the sparse methods (see Appendix \ref{appendix:tables:eu}, Tables \ref{es:p51g}, \ref{es:p6}, \ref{fr:p3}, \ref{fr:p51g} for Spain GFCF and Exports, France PFCE and GFCF). 

In fact, for Italy PFCE variable, both the sparse and dense methods were hardly outperforming the ARIMA models (see Appendix \ref{appendix:tables:eu}, Table \ref{it:p3}) when nowcasting. This result naturally follows from the scarce availability of data and the short historical periods for the available explanatory variables. 

Some variants of the proposed LASSO-PC modification were able to generate the most accurate nowcasts for France Exports, Italy Imports (see Appendix \ref{appendix:tables:eu}, Tables \ref{fr:p6}, \ref{it:p7}), and the most accurate 1- or 2-quarter forecasts for Spain (GFCF, Imports), Germany (PFCE, GFCF, Exports), France (GFCF, Exports, Imports), Italy (PFCE, GFCF, Imports) (see Appendix \ref{appendix:tables:eu}, Tables \ref{es:p51g},\ref{es:p7}, \ref{de:p3}, \ref{de:p51g}, \ref{de:p6}, \ref{fr:p51g}, \ref{fr:p6}, \ref{fr:p7}, \ref{it:p3}, \ref{it:p51g}, \ref{it:p7}). While some of the gains are less evident according the estimated statistical significance using the GW test, the results extend the findings from the US forecasting results. Noteworthy, in many cases the nowcasting performance for the DF and BNDF models were significantly improved from applying commonly used monthly aggregation strategies and the use of bridge equations for the nowcast estimation, as described in \cite{BANBURA2013195}, which were not used for the sparse methods. On the other hand, when comparing the 1- and 2- quarter forecasting performance, the effects of such nowcasting solutions are diminished and a clearer comparison can be made.

Comparing the performance by the forecasted GDP components, we conclude that for all four countries, the PCA modification consistently showed improvements for the GFCF. However, for the remaining variables, the performance gains were less evident and less consistent. These results confirm the ideas discussed in Sections \ref{chapter:PR} and \ref{expW:gfcf} for the US data and provides further evidence that the structural complexity of the GFCF component and the lack of ``hard'' leading indicators translates well within Germany, Spain, France, and Italy. Hence, in such cases, we can expect tangible gains from applying LASSO-PC against traditional sparse methods. 

Finally, in addition to inspecting only the best performing models, we examine the median overall performance rankings of every model in the ``horse race''. Table \ref{tables:rankings} presents the results, where the models are ordered by their median rankings throughout the ``horse race''. Median rankings can demonstrate the overall stability of the models, capturing the relative model performance under varying data quality and availability. In this case, it follows that the LP and LPX models are able to consistently generate adequate nowcasts and 1- and 2-quarter forecasts, while the remaining models show higher variability and higher probability of producing relatively poor forecasts. 

\setlongtables{\scriptsize
\begin{longtable}{llllll}\caption{Median model rankings during the pseudo-real experiments for Euro Area GDP components over the full period of 2005-2019. The values are sorted by the 1-quarter forecast performance in the ascending order. \label{tables:rankings}} \tabularnewline*

			Model & Nowcast & Forecast-1Q & Forecast-2Q &  \\ \midrule
			LPX   & 7         & 5          & 5.5         &  \\
			LP    & 5.5       & 6          & 7           &  \\
			L0L1  & 9         & 7.5        & 12          &  \\
			L     & 8         & 8          & 7           &  \\
			SqL   & 8         & 8          & 8.5         &  \\
			SqP   & 8.5       & 8          & 8           &  \\
			AdP   & 8         & 8.5        & 7.5         &  \\
			RL    & 10        & 8.5        & 6.5         &  \\
			L0L2p & 10        & 9          & 10.5        &  \\
			AdL   & 7.5       & 9.5        & 9.5         &  \\
			L0L1p & 11.5      & 10         & 12          &  \\
			DF    & 11        & 11         & 10.5        &  \\
			BNDF  & 6.5       & 11.5       & 12.5        &  \\
			L0L2  & 13        & 11.5       & 8.5         &  \\
			AdRL  & 8.5       & 12         & 9.5         &  \\
			L0    & 16        & 15         & 16          &  \\
			L0p   & 16        & 16         & 15          &  \\
			SW    & 17.5      & 17         & 8.5         &  \\
			ARMA  & 16        & 18         & 18          & \\ \bottomrule

\end{longtable}}

\section{Conclusions and final remarks}

Short-term forecasting of quarterly components of the GDP relies on the availability of timely monthly information. In this paper, we studied the forecasting accuracy of the LASSO and its widespread modifications, together with our proposed approach of combining LASSO with the method of principal components. These approaches assume a sparse structure of the available information set required for adequate modelling. Thus, the methods can distinguish and estimate the underlying explanatory variables for the nowcasting, short-run forecasting problem. Conducting a pseudo-real-time forecasting exercise, we analysed forecasting performance of the approaches, from which three main results emerge:

First, all LASSO methods show acceptable forecasting performance, outperforming the benchmark ARMA and factor models. The advantages of including additional explanatory monthly information are substantial during the crisis period of 2008Q1-2010Q4, where both the nowcasts and 1- and 2-quarter forecasts in most cases provide more accurate results than the benchmark model. Furthermore, the number of variables selected by the methods mostly was relatively small, suggesting that the sparseness assumption for the data generating process holds.

Second, in most of the cases the modifications of LASSO, discussed in this paper, are able to improve the forecasting accuracy of the LASSO, suggesting not only theoretical but also practical usefulness of looking into the modifications of the classic LASSO method. 

Third, while the LASSO is fit for generating adequate forecasts for different macroeconomic data, our suggested modification by combining the methodologies of LASSO and the principal components show additional gains in forecast accuracy. Namely, we found evidence that the proposed combination often generates more accurate forecasts than the Adaptive LASSO or the Relaxed LASSO, which already are Oracle properties having modifications of the original LASSO. Therefore, we expect further gains in forecasting accuracy with additional work on these methods. On the other hand, the discussed methods never find non-linearities if they are not in the initial search space. More time consuming, yet interesting extension, in this regard, would be to consider second/third order interaction terms between the variables or their power transformations, which might result in further improvement of forecasting performance.

The results presented in this paper might be further improved working in two specific directions. First, the usage of default weights for the Adaptive LASSO often showed better pseudo-real-time forecasting performance than the standard LASSO and benchmark models. We suppose that the choice of the weights could be optimized or cross-validated to better deal with the high-dimensionality problem. Second, while combining the method of principal components with the LASSO, only the standard estimation procedure of the principal components was considered. However, the PCA part of the proposed in the paper approach deemed to benefit further from tailoring the angles or scale of the rotation matrix and show additional gains in forecasting accuracy.

\bigskip
{\bf Acknowledgements}. The authors would like to thank the anonymous Referee for his/her very constructive and detailed comments and suggestions on the first version of the manuscript. Remigijus Leipus acknowledges the support from the grant No.\ S-MIP-20-16 from the Research Council of Lithuania.

\bibliography{reference}

\begin{appendices}

	\section{Data preparation}
	\label{appendix:data}
	 Due to the nature of $p \gg n$ problem, it is relatively easy for automatic selection algorithms such as LASSO to find the spurious signal in the noise. Applying linear regression, we could discard such noise variables through cross-validation. However, the cross-validation becomes problematic when shrinkage is applied, since the spurious effects get smoothed out in out-of-sample forecasts. For this reason, we use a strict set of rules for variable prescreening.
	 
	 First, by performing the \textit{Kwiatkowski–Phillips–Schmidt–Shin} ({KPSS}) test, the stationarity of the time series was tested (with 5\% significance). Second, the test for the unit-roots was performed by using the \textit{Augmented Dickey-Fuller} ({ADF}) test, accounting for the probable deterministic part of the data. Additionally, since the test statistic of the ADF test is the estimated value of the $t-$statistics from an auxiliary regression model, its resulting residuals were also inspected for the possible presence of heteroscedasticity. If we could not reject the heteroscedasticity hypothesis, according to \textit{Breusch-Pagan} {(BP)} test with 5\% significance level, the estimated value of the $t$-statistics might be biased. Therefore, in such cases, we additionally perform nonparametric \textit{Philips-Perron} {(PP)} unit-root test, which can correct the possibly incorrect results of the ADF test by bootstrapping the critical values of the test statistic. 
	 
	In all cases, we determined the number of lags used in the auxiliary regressions by minimizing the \textit{Akaike} information criterion. The time series was found as statistically significantly non-stationary if either KPSS or unit-root tests suggested non-stationarity with 5\% significance. In such case, the series was transformed by differencing, after which the whole stationarity testing procedure was iterated until the final series was found as significantly stationary\footnote{It can be noted that no series required more than 2 differences to achieve stationarity.}.
	 
Since most of the macroeconomic indicators used in the paper follow multiplicative processes, it is instructive to apply logarithmic transformation where relevant. By using the logarithmic transformation, we transform the underlying multiplicative processes into additive, removing most of the explosive effects and normalizing the variability of the data. Finally, log-transformation does not alter the interpretation of the variables a lot, because the differences of log-transformed data approximate the percentage growths of the original data.

We test the relevance of logarithmic transformation by estimating and interpreting the parameter of Box-Cox transformation (\cite{10.2307/2984418}). A logarithmic transformation was applied if the Box-Cox parameter estimate of the analyzed indicator is smaller than 0.8. The higher threshold value was chosen to safeguard against possible structural breaks in the trend of the data. Transformations of $\frac{x^q-1}{q}$, $q \in (0,1)$ were not used in this paper since the focus is on extracting and distinguishing the multiplicative effects if such were present, instead of just normalizing the data. 
	 
	 Finally, some of the available data has relatively large spikes (outliers) at particular periods, with comparably small volatility during the other remaining periods. Therefore, such a variable may be included in the final model not as an explanatory variable, but rather as a dummy variable, helping the model fitting some of the sudden shocks in the data, but providing no additional information to the forecasts\footnote{It is even worse if the sudden shock is relatively recent, since it may strongly affect the individual forecasts of such a series.}. Therefore, we applied an additional heuristic rule to filter such variables from the final dataset: the variable was not included in the final dataset if the ratio of maximum to the average value, when adjusted by standard deviation, was higher than 10. We found that the inclusion of such variables to the final dataset resulted in worsened forecasting accuracy, especially during the crisis periods, when they were included in the models as dummy variables to explain the sudden shock. 
	 
	 \section{Main variables used in the modelling}

	 \setlongtables{\scriptsize
\begin{longtable}{p{3.5cm}llp{9.5cm}}\caption{US GFCF (expanding window): acronyms and full names of all of the variables, used in presenting the modelling results. The acronyms correspond to those originally used in the FRED's database.}\label{variables:expanding} \tabularnewline
	 		\hline\hline
	 		\multicolumn{1}{c}{Acronym}&\multicolumn{1}{c}{Transf.}&\multicolumn{1}{c}{Lag}&\multicolumn{1}{c}{Description}\tabularnewline
	 		\hline
	 		\endfirsthead\caption[]{\em (continued)} \tabularnewline
	 		\hline
	 		\multicolumn{1}{c}{Acronym}&\multicolumn{1}{c}{Transf.}&\multicolumn{1}{c}{Lag}&\multicolumn{1}{c}{Description}\tabularnewline
	 		\hline
	 		\endhead
	 		\hline
	 		\endfoot
	 		ANXAVS&$\Delta^{1} \ $&1&Value of Manufacturers' Shipments for Capital Goods: Nondefense Capital Goods Excluding Aircraft Industries\tabularnewline
	 		CACONS&$\Delta^{1} \ln $&1&All Employees: Construction in California\tabularnewline
	 		CES6562000101&$\Delta^{1} \ $&1&All Employees: Education and Health Services: Health Care\tabularnewline
	 		CUUR0000SETC&$\Delta^{1} \ln $&1&Consumer Price Index for All Urban Consumers: Motor vehicle parts and equipment\tabularnewline
	 		IAWARR1URN&$\Delta^{1} \ln $&2&Unemployment Rate in Warren County, IA\tabularnewline
	 		IMPMX&$\Delta^{1} \ln $&2&U.S. Imports of Goods from Mexico, Customs Basis\tabularnewline
	 		IPDMAN&$\Delta^{1} \ln $&1&Industrial Production: Durable Manufacturing (NAICS)\tabularnewline
	 		KYMADI0URN&$\Delta^{1} \ln $&2&Unemployment Rate in Madison County, KY\tabularnewline
\hline
	 \end{longtable}}

	 \setlongtables{\scriptsize
	 	\begin{longtable}{p{3.5cm}llp{9.5cm}}\caption{US GFCF (rolling window): acronyms and full names of all of the variables, used in presenting the modelling results. The acronyms correspond to those originally used in the FRED's database.}\label{variables:rolling} \tabularnewline
	 		\hline\hline
	 		\multicolumn{1}{c}{Acronym}&\multicolumn{1}{c}{Transf.}&\multicolumn{1}{c}{Lag}&\multicolumn{1}{c}{Description}\tabularnewline
	 		\hline
	 		\endfirsthead\caption[]{\em (continued)} \tabularnewline
	 		\hline
	 		\multicolumn{1}{c}{Acronym}&\multicolumn{1}{c}{Transf.}&\multicolumn{1}{c}{Lag}&\multicolumn{1}{c}{Description}\tabularnewline
	 		\hline
	 		\endhead
	 		\hline
	 		\endfoot
	 		CES2023800001&$\Delta^{1} \ln $&1&All Employees: Construction: Specialty Trade Contractors\tabularnewline
	 		CES6562000101&$\Delta^{1} \ $&1&All Employees: Education and Health Services: Health Care\tabularnewline
	 		INDI918NAN&$\Delta^{1} \ $&1&All Employees: Total Nonfarm in Indianapolis-Carmel-Anderson, IN (MSA)\tabularnewline
	 		IPG326S&$\Delta^{1} \ $&1&Industrial Production: Nondurable manufacturing: Plastics and rubber products\tabularnewline
	 		IR3TBB01NZM156N&$\Delta^{1} \ln $&1&3-Month or 90-day Rates and Yields: Bank Bills for New Zealand©\tabularnewline
	 		KYNAN&$\Delta^{1} \ $&1&All Employees: Total Nonfarm in Kentucky\tabularnewline
	 		LREM25MAUSM156N&$\Delta^{1} \ $&1&Employment Rate: Aged 25-54: Males for the United States©\tabularnewline
	 		MINN427NAN&$\Delta^{1} \ $&1&All Employees: Total Nonfarm in Minneapolis-St. Paul-Bloomington, MN-WI (MSA)\tabularnewline
\hline
	 \end{longtable}}
 
 \setlongtables{\scriptsize
 	\begin{longtable}{p{3.5cm}llp{9.5cm}}\caption{US PFCE (rolling window): acronyms and full names of all of the variables, used in presenting the modelling results. The acronyms correspond to those originally used in the FRED's database.}\label{variables:rolling:pfce} \tabularnewline
 		\hline\hline
 		\multicolumn{1}{c}{Acronym}&\multicolumn{1}{c}{Transf.}&\multicolumn{1}{c}{Lag}&\multicolumn{1}{c}{Description}\tabularnewline
 		\hline
 		\endfirsthead\caption[]{\em (continued)} \tabularnewline
 		\hline
 		\multicolumn{1}{c}{Acronym}&\multicolumn{1}{c}{Transf.}&\multicolumn{1}{c}{Lag}&\multicolumn{1}{c}{Description}\tabularnewline
 		\hline
 		\endhead
 		\hline
 		\endfoot
 		ACDGNO&$\Delta^{1} \ $&2&Value of Manufacturers' New Orders for Consumer Goods: Consumer Durable Goods Industries\tabularnewline
 		CABPPRIV&$\Delta^{1} \ln $&1&New Private Housing Units Authorized by Building Permits for California\tabularnewline
 		CES2023700001&$\Delta^{1} \ln $&1&All Employees: Construction: Heavy and Civil Engineering Construction\tabularnewline
 		CEU4244100001&$\Delta^{1} \ $&1&All Employees: Retail Trade: Motor Vehicle and Parts Dealers\tabularnewline
 		CEU5552000001&$\Delta^{1} \ $&1&All Employees: Financial Activities: Finance and Insurance\tabularnewline
 		CEU6562000101&$\Delta^{1} \ $&1&All Employees: Education and Health Services: Health Care\tabularnewline
 		CMRMT&$\Delta^{1} \ln $&2&Real Manufacturing and Trade Industries Sales\tabularnewline
 		CTUR&$\Delta^{1} \ln $&1&Unemployment Rate in Connecticut\tabularnewline
\hline
 \end{longtable}}
\setlongtables{\scriptsize
	\begin{longtable}{p{3.5cm}llp{9.5cm}}\caption{US Exports (rolling window): acronyms and full names of all of the variables, used in presenting the modelling results. The acronyms correspond to those originally used in the FRED's database.}\label{variables:rolling:exp} \tabularnewline
		\hline\hline
		\multicolumn{1}{c}{Acronym}&\multicolumn{1}{c}{Transf.}&\multicolumn{1}{c}{Lag}&\multicolumn{1}{c}{Description}\tabularnewline
		\hline
		\endfirsthead\caption[]{\em (continued)} \tabularnewline
		\hline
		\multicolumn{1}{c}{Acronym}&\multicolumn{1}{c}{Transf.}&\multicolumn{1}{c}{Lag}&\multicolumn{1}{c}{Description}\tabularnewline
		\hline
		\endhead
		\hline
		\endfoot
		A36CTI&$\Delta^{1} \ln $&2&Value of Manufacturers' Total Inventories for Durable Goods Industries: Transportation Equipment: Heavy Duty Trucks\tabularnewline
		ANXAVS&$\Delta^{1} \ $&1&Value of Manufacturers' Shipments for Capital Goods: Nondefense Capital Goods Excluding Aircraft Industries\tabularnewline
		AUENSA&$\Delta^{1} \ln $&2&Auto Exports\tabularnewline
		BOPTEXP&$\Delta^{1} \ln $&2&Exports of Goods and Services, Balance of Payments Basis\tabularnewline
		CODENV5URN&$\Delta^{1} \ln $&2&Unemployment Rate in Denver County/city, CO\tabularnewline
		EXBZUS&$\Delta^{1} \ln $&1&Brazil / U.S. Foreign Exchange Rate\tabularnewline
		IPG3344S&$\Delta^{2} \ln $&1&Industrial Production: Durable Goods: Semiconductor and other electronic component\tabularnewline
		IPHITEK2N&$\Delta^{2} \ln $&1&Industrial Production: Computers, communications equipment, and semiconductors\tabularnewline
\hline
\end{longtable}}
\setlongtables{\scriptsize
	\begin{longtable}{p{3.5cm}llp{9.5cm}}\caption{US Imports (rolling window): acronyms and full names of all of the variables, used in presenting the modelling results. The acronyms correspond to those originally used in the FRED's database.}\label{variables:rolling:imp} \tabularnewline
		\hline\hline
		\multicolumn{1}{c}{Acronym}&\multicolumn{1}{c}{Transf.}&\multicolumn{1}{c}{Lag}&\multicolumn{1}{c}{Description}\tabularnewline
		\hline
		\endfirsthead\caption[]{\em (continued)} \tabularnewline
		\hline
		\multicolumn{1}{c}{Acronym}&\multicolumn{1}{c}{Transf.}&\multicolumn{1}{c}{Lag}&\multicolumn{1}{c}{Description}\tabularnewline
		\hline
		\endhead
		\hline
		\endfoot
		A31SNO&$\Delta^{1} \ln $&1&Value of Manufacturers' New Orders for Durable Goods Industries: Primary Metals\tabularnewline
		BSCICP03USM665S& &1&Business Tendency Surveys for Manufacturing: Confidence Indicators: Composite Indicators: OECD Indicator for the United States©\tabularnewline
		CANPROINDMISMEI&$\Delta^{1} \ $&3&Production of Total Industry in Canada©\tabularnewline
		CES2023610001&$\Delta^{1} \ln $&1&All Employees: Construction: Residential Building\tabularnewline
		CEU4348800001&$\Delta^{1} \ $&1&All Employees: Transportation and Warehousing: Support Activities for Transportation\tabularnewline
		CILLCBM027NBOG&$\Delta^{1} \ln $&1&Commercial and Industrial Loans, Large Domestically Chartered Commercial Banks\tabularnewline
		COMPU1UNSA&$\Delta^{1} \ln $&1&New Privately-Owned Housing Units Completed: 1-Unit Structures\tabularnewline
		ESPPROINDMISMEI&$\Delta^{1} \ln $&2&Production of Total Industry in Spain©\tabularnewline
\hline
\end{longtable}}

	 \section{Tables}
	 \label{appendix:tables}
	 \subsection{US}
	 
	 \setlongtables{\scriptsize
	 	\begin{longtable}{lccccccccc|ccccccccccccc} \caption{RMSE of models forecasts during rolling window pseudo-real-time experiments for US PFCE, Exports and Imports; here the bolded values are the smallest ones for every row during the full time period of 2005Q1-2018Q4 \label{table:final} } \tabularnewline \hline\hline
	 		\multicolumn{1}{l}{\bfseries \rule{0pt}{2.5ex}  \vspace{0.02 in}  }&\multicolumn{1}{c}{\bfseries LP}&\multicolumn{1}{c}{\bfseries LPX}&\multicolumn{1}{c}{\bfseries RL}&\multicolumn{1}{c}{\bfseries SqL}&\multicolumn{1}{c}{\bfseries SqP}&\multicolumn{1}{c}{\bfseries AdP}&\multicolumn{1}{c}{\bfseries AdL}&\multicolumn{1}{c}{\bfseries AdRL}&\multicolumn{1}{c}{\bfseries L}&\multicolumn{1}{c}{\bfseries SW}&\multicolumn{1}{c}{\bfseries BNDF}&\multicolumn{1}{c}{\bfseries DF}&\multicolumn{1}{c}{\bfseries ARMA}\tabularnewline*
	 		\hline
	 		\endhead
	 		\hline
	 		\endfoot
	 		\rule{0pt}{2.5ex} 
	 		{\bfseries PFCE:}&&&&&&&&&&&&&\tabularnewline*
	 		\rule{0pt}{2.5ex} 
	 		~\textcolor{black}{Backcast}&\textcolor{black}{0.08}&\textcolor{black}{0.07}&\textcolor{black}{0.07}&\textcolor{black}{0.14}&\textcolor{black}{0.10}&\textcolor{black}{\textbf{0.03}}&\textcolor{black}{0.04}&\textcolor{black}{0.03}&\textcolor{black}{0.11}&\textcolor{black}{0.44}&\textcolor{black}{0.27}&\textcolor{black}{0.37}&\textcolor{black}{0.36}\tabularnewline
	 		
	 		~~\textcolor{black}{Nowcast}&\textcolor{black}{0.29}&\textcolor{black}{0.29}&\textcolor{black}{0.29}&\textcolor{black}{\textbf{0.28}}&\textcolor{black}{0.29}&\textcolor{black}{0.29}&\textcolor{black}{0.29}&\textcolor{black}{0.29}&\textcolor{black}{0.28}&\textcolor{black}{0.53}&\textcolor{black}{0.30}&\textcolor{black}{0.48}&\textcolor{black}{0.38}\tabularnewline
	 		
	 		~~\textcolor{black}{Forecast-1Q}&\textcolor{black}{0.38}&\textcolor{black}{0.38}&\textcolor{black}{0.38}&\textcolor{black}{0.40}&\textcolor{black}{\textbf{0.37}}&\textcolor{black}{0.39}&\textcolor{black}{0.39}&\textcolor{black}{0.39}&\textcolor{black}{0.39}&\textcolor{black}{0.57}&\textcolor{black}{0.37}&\textcolor{black}{0.54}&\textcolor{black}{0.42}\tabularnewline
	 		
	 		~~\textcolor{black}{Forecast-2Q}&\textcolor{black}{\textbf{0.43}}&\textcolor{black}{0.44}&\textcolor{black}{0.44}&\textcolor{black}{0.46}&\textcolor{black}{0.44}&\textcolor{black}{0.44}&\textcolor{black}{0.44}&\textcolor{black}{0.44}&\textcolor{black}{0.43}&\textcolor{black}{0.69}&\textcolor{black}{0.46}&\textcolor{black}{0.55}&\textcolor{black}{0.47}\tabularnewline
	 		\rule{0pt}{2.5ex} 
	 		{\bfseries Exports:}&&&&&&&&&&&&&\tabularnewline
\end{longtable}}
	 
	 \setlongtables{\scriptsize
	 	\begin{longtable}{lcccccc|ccccccccc}\caption{RMSE of models forecasts during rolling window pseudo-real-time experiments for US PFCE, Exports and Imports; here the bolded values are the smallest ones for every row during the full time period of 2005Q1-2018Q4 \label{table:final2} } \tabularnewline\hline\hline
	 		\multicolumn{1}{l}{\bfseries \rule{0pt}{2.5ex}  \vspace{0.02 in} }&\multicolumn{1}{c}{\bfseries L0L1}&\multicolumn{1}{c}{\bfseries L0}&\multicolumn{1}{c}{\bfseries L0L2}&\multicolumn{1}{c}{\bfseries L0L1p}&\multicolumn{1}{c}{\bfseries L0p}&\multicolumn{1}{c}{\bfseries L0L2p}&\multicolumn{1}{c}{\bfseries SW}&\multicolumn{1}{c}{\bfseries BNDF}&\multicolumn{1}{c}{\bfseries DF}&\multicolumn{1}{c}{\bfseries ARMA}\tabularnewline*
	 		\endhead
	 		\hline
	 		\endfoot
	 		\hline
	 		\rule{0pt}{2.5ex} 
	 		{\bfseries PFCE:}&&&&&&&&&&\tabularnewline*
	 		\rule{0pt}{2.5ex} 
	 		~\textcolor{black}{Backcast}&\textcolor{black}{0.18}&\textcolor{black}{0.25}&\textcolor{black}{0.19}&\textcolor{black}{\textbf{0.17}}&\textcolor{black}{0.25}&\textcolor{black}{0.18}&\textcolor{black}{0.44}&\textcolor{black}{0.27}&\textcolor{black}{0.37}&\textcolor{black}{0.36}\tabularnewline
	 		
	 		~~\textcolor{black}{Nowcast}&\textcolor{black}{0.29}&\textcolor{black}{0.43}&\textcolor{black}{0.32}&\textcolor{black}{\textbf{0.27}}&\textcolor{black}{0.42}&\textcolor{black}{0.30}&\textcolor{black}{0.53}&\textcolor{black}{0.30}&\textcolor{black}{0.48}&\textcolor{black}{0.38}\tabularnewline
	 		
	 		~~\textcolor{black}{Forecast-1Q}&\textcolor{black}{0.44}&\textcolor{black}{0.50}&\textcolor{black}{0.43}&\textcolor{black}{0.43}&\textcolor{black}{0.49}&\textcolor{black}{0.42}&\textcolor{black}{0.57}&\textcolor{black}{\textbf{0.37}}&\textcolor{black}{0.54}&\textcolor{black}{0.42}\tabularnewline
	 		
	 		~~\textcolor{black}{Forecast-2Q}&\textcolor{black}{\textbf{0.46}}&\textcolor{black}{0.54}&\textcolor{black}{0.47}&\textcolor{black}{0.46}&\textcolor{black}{0.53}&\textcolor{black}{0.48}&\textcolor{black}{0.69}&\textcolor{black}{0.46}&\textcolor{black}{0.55}&\textcolor{black}{0.47}\tabularnewline
	 		\rule{0pt}{2.5ex} 
	 		{\bfseries Exports:}&&&&&&&&&&\tabularnewline*
	 		\rule{0pt}{2.5ex} 
	 		~\textcolor{black}{Backcast}&\textcolor{black}{0.67}&\textcolor{black}{0.99}&\textcolor{black}{0.97}&\textcolor{black}{\textbf{0.55}}&\textcolor{black}{0.98}&\textcolor{black}{0.86}&\textcolor{black}{1.28}&\textcolor{black}{1.00}&\textcolor{black}{1.35}&\textcolor{black}{1.66}\tabularnewline
\hline
	 \end{longtable}}

	 \setlongtables{\scriptsize
	 	\begin{longtable}{lclllclllclllclll}\caption{Relative RMSE of model forecasts during the rolling window pseudo-real-time experiments for US PFCE against four benchmarks: ARMA, BNDF, DF and SW models. The labels N, Q1 and Q2 denotes the forecast horizons -- Nowcast, Forecast-1Q and Forecast-2Q, respectively. The bolded values correspond to the greatest accuracy within the forecast horizon, while the asterix denotes significant performance improvement as suggested by GW test with 5\% significance. \label{pfce:tables}} \tabularnewline
	 		\hline\hline
	 		\multicolumn{1}{c}{\bfseries Model}&\multicolumn{1}{c}{\bfseries \rule{0pt}{2.5ex}  \vspace{0.02 in} }&\multicolumn{3}{c}{\bfseries ARMA}&\multicolumn{1}{c}{\bfseries }&\multicolumn{3}{c}{\bfseries BNDF}&\multicolumn{1}{c}{\bfseries }&\multicolumn{3}{c}{\bfseries DF}&\multicolumn{1}{c}{\bfseries }&\multicolumn{3}{c}{\bfseries SW}\tabularnewline
	 		\cline{3-5} \cline{7-9} \cline{11-13} \cline{15-17}
	 		\multicolumn{1}{c}{\rule{0pt}{2.5ex}  \vspace{0.02 in}}&\multicolumn{1}{c}{}&\multicolumn{1}{c}{\textbf{N}}&\multicolumn{1}{c}{\textbf{Q1}}&\multicolumn{1}{c}{\textbf{Q2}}&\multicolumn{1}{c}{}&\multicolumn{1}{c}{\textbf{N}}&\multicolumn{1}{c}{\textbf{Q1}}&\multicolumn{1}{c}{\textbf{Q2}}&\multicolumn{1}{c}{}&\multicolumn{1}{c}{\textbf{N}}&\multicolumn{1}{c}{\textbf{Q1}}&\multicolumn{1}{c}{\textbf{Q2}}&\multicolumn{1}{c}{}&\multicolumn{1}{c}{\textbf{N}}&\multicolumn{1}{c}{\textbf{Q1}}&\multicolumn{1}{c}{\textbf{Q2}}\tabularnewline
	 		\hline
	 		\endfirsthead\caption[]{\em (continued)} \tabularnewline
	 		\hline
	 		\multicolumn{1}{c}{\bfseries Model}&\multicolumn{1}{c}{\bfseries \rule{0pt}{2.5ex}  \vspace{0.02 in}}&\multicolumn{3}{c}{\bfseries ARMA}&\multicolumn{1}{c}{\bfseries }&\multicolumn{3}{c}{\bfseries BNDF}&\multicolumn{1}{c}{\bfseries }&\multicolumn{3}{c}{\bfseries DF}&\multicolumn{1}{c}{\bfseries }&\multicolumn{3}{c}{\bfseries SW}\tabularnewline
	 		\cline{3-5} \cline{7-9} \cline{11-13} \cline{15-17}
	 		\multicolumn{1}{c}{\rule{0pt}{2.5ex}  \vspace{0.02 in}}&\multicolumn{1}{c}{}&\multicolumn{1}{c}{\textbf{N}}&\multicolumn{1}{c}{\textbf{Q1}}&\multicolumn{1}{c}{\textbf{Q2}}&\multicolumn{1}{c}{}&\multicolumn{1}{c}{\textbf{N}}&\multicolumn{1}{c}{\textbf{Q1}}&\multicolumn{1}{c}{\textbf{Q2}}&\multicolumn{1}{c}{}&\multicolumn{1}{c}{\textbf{N}}&\multicolumn{1}{c}{\textbf{Q1}}&\multicolumn{1}{c}{\textbf{Q2}}&\multicolumn{1}{c}{}&\multicolumn{1}{c}{\textbf{N}}&\multicolumn{1}{c}{\textbf{Q1}}&\multicolumn{1}{c}{\textbf{Q2}}\tabularnewline
	 		\hline
	 		\endhead
	 		\hline
	 		\endfoot
	 		AdL&\rule{0pt}{2.5ex}&0.77&0.92&0.94&&0.97&1.05&0.96&&0.61*&0.72*&0.81&&0.54*&0.68*&0.64*\tabularnewline
	 		AdP&&0.76&0.92&0.92&&0.96&1.05&0.94&&0.60*&0.72*&0.80&&0.54*&0.68*&0.63*\tabularnewline
	 		AdRL&&0.77&0.92&0.93&&0.98&1.05&0.95&&0.61*&0.72*&0.80&&0.55*&0.68*&0.63*\tabularnewline
	 		L&&0.75&0.93&0.92&&0.95&1.05&0.94&&0.60*&0.72*&0.79*&&0.53*&0.68*&0.63*\tabularnewline
	 		L0&&1.13&1.19&1.14&&1.43&1.36&1.16&&0.89&0.93&0.98&&0.80&0.88&0.78*\tabularnewline
	 		L0L1&&0.76&1.04&0.98&&0.96&1.19&1.00&&0.60*&0.81&0.84*&&0.54*&0.77&0.67*\tabularnewline
	 		L0L1p&&\textbf{0.72*}&1.03&0.97&&\textbf{0.91}&1.17&0.99&&\textbf{0.57*}&0.80&0.83&&\textbf{0.51*}&0.76&0.66*\tabularnewline
	 		L0L2&&0.85&1.02&0.99&&1.07&1.16&1.02&&0.67*&0.80&0.86&&0.60*&0.75&0.68*\tabularnewline
\hline
	 \end{longtable}}

	 \setlongtables{\scriptsize
	 	\begin{longtable}{lclllclllclllclll}\caption{Relative RMSE of model forecasts during the rolling window pseudo-real-time experiments for US Exports against four benchmarks: ARMA, BNDF, DF and SW models. The labels N, Q1 and Q2 denotes the forecast horizons -- Nowcast, Forecast-1Q and Forecast-2Q, respectively. The bolded values correspond to the greatest accuracy within the forecast horizon, while the asterix denotes significant performance improvement as suggested by GW test with 5\% significance. \label{exp:tables}} \tabularnewline
	 		\hline\hline
	 		\multicolumn{1}{c}{\bfseries Model}&\multicolumn{1}{c}{\bfseries \rule{0pt}{2.5ex}  \vspace{0.02 in} }&\multicolumn{3}{c}{\bfseries ARMA}&\multicolumn{1}{c}{\bfseries }&\multicolumn{3}{c}{\bfseries BNDF}&\multicolumn{1}{c}{\bfseries }&\multicolumn{3}{c}{\bfseries DF}&\multicolumn{1}{c}{\bfseries }&\multicolumn{3}{c}{\bfseries SW}\tabularnewline
	 		\cline{3-5} \cline{7-9} \cline{11-13} \cline{15-17}
	 		\multicolumn{1}{c}{ \rule{0pt}{2.5ex}  \vspace{0.02 in} }&\multicolumn{1}{c}{}&\multicolumn{1}{c}{\textbf{N}}&\multicolumn{1}{c}{\textbf{Q1}}&\multicolumn{1}{c}{\textbf{Q2}}&\multicolumn{1}{c}{}&\multicolumn{1}{c}{\textbf{N}}&\multicolumn{1}{c}{\textbf{Q1}}&\multicolumn{1}{c}{\textbf{Q2}}&\multicolumn{1}{c}{}&\multicolumn{1}{c}{\textbf{N}}&\multicolumn{1}{c}{\textbf{Q1}}&\multicolumn{1}{c}{\textbf{Q2}}&\multicolumn{1}{c}{}&\multicolumn{1}{c}{\textbf{N}}&\multicolumn{1}{c}{\textbf{Q1}}&\multicolumn{1}{c}{\textbf{Q2}}\tabularnewline
	 		\hline
	 		\endfirsthead\caption[]{\em (continued)} \tabularnewline
	 		\hline
	 		\multicolumn{1}{c}{\bfseries Model}&\multicolumn{1}{c}{\bfseries \rule{0pt}{2.5ex}  \vspace{0.02 in}}&\multicolumn{3}{c}{\bfseries ARMA}&\multicolumn{1}{c}{\bfseries }&\multicolumn{3}{c}{\bfseries BNDF}&\multicolumn{1}{c}{\bfseries }&\multicolumn{3}{c}{\bfseries DF}&\multicolumn{1}{c}{\bfseries }&\multicolumn{3}{c}{\bfseries SW}\tabularnewline
	 		\cline{3-5} \cline{7-9} \cline{11-13} \cline{15-17}
	 		\multicolumn{1}{c}{}&\multicolumn{1}{c}{}&\multicolumn{1}{c}{\textbf{N}}&\multicolumn{1}{c}{\textbf{Q1}}&\multicolumn{1}{c}{\textbf{Q2}}&\multicolumn{1}{c}{}&\multicolumn{1}{c}{\textbf{N}}&\multicolumn{1}{c}{\textbf{Q1}}&\multicolumn{1}{c}{\textbf{Q2}}&\multicolumn{1}{c}{}&\multicolumn{1}{c}{\textbf{N}}&\multicolumn{1}{c}{\textbf{Q1}}&\multicolumn{1}{c}{\textbf{Q2}}&\multicolumn{1}{c}{}&\multicolumn{1}{c}{\textbf{N}}&\multicolumn{1}{c}{\textbf{Q1}}&\multicolumn{1}{c}{\textbf{Q2}}\tabularnewline
	 		\hline
	 		\endhead
	 		\hline
	 		\endfoot
	 		AdL&\rule{0pt}{2.5ex}&0.68*&0.96&1.01&&1.07&1.11&0.98&&0.82&1.04&1.09&&0.86&0.99&0.94\tabularnewline
	 		AdP&&0.69*&0.91&1.00&&1.09&1.06&0.97&&0.83&0.99&1.07&&0.88&0.94&0.93\tabularnewline
	 		AdRL&&0.69*&0.94&1.01&&1.09&1.10&0.99&&0.83&1.03&1.09&&0.88&0.97&0.94\tabularnewline
	 		L&&0.66*&0.91&\textbf{0.97}&&1.03&1.06&\textbf{0.95}&&0.79*&0.99&\textbf{1.05}&&0.83&0.94&\textbf{0.90}\tabularnewline
	 		L0&&0.73&1.00&0.98&&1.15&1.16&0.96&&0.88&1.09*&1.06&&0.93&1.03&0.92\tabularnewline
	 		L0L1&&0.66*&1.01&0.99&&1.04&1.17&0.97&&0.79&1.10*&1.07*&&0.84&1.04*&0.92\tabularnewline
	 		L0L1p&&0.68*&1.02&0.99&&1.06&1.19&0.97&&0.81&1.11*&1.07*&&0.86&1.05*&0.92\tabularnewline
	 		L0L2&&0.73*&1.04&1.00&&1.15&1.21&0.98&&0.88&1.13*&1.08&&0.93&1.07*&0.93\tabularnewline
\hline
	 \end{longtable}}
	 
	 \setlongtables{\scriptsize
	 	\begin{longtable}{lclllclllclllclll}\caption{Relative RMSE of model forecasts during the rolling window pseudo-real-time experiments for US Imports against four benchmarks: ARMA, BNDF, DF and SW models. The labels N, Q1 and Q2 denotes the forecast horizons -- Nowcast, Forecast-1Q and Forecast-2Q, respectively. The bolded values correspond to the greatest accuracy within the forecast horizon, while the asterix denotes significant performance improvement as suggested by GW test with 5\% significance. \label{imp:tables}} \tabularnewline
	 		\hline\hline
	 		\multicolumn{1}{c}{\bfseries Model}&\multicolumn{1}{c}{\bfseries \rule{0pt}{2.5ex}  \vspace{0.02 in} }&\multicolumn{3}{c}{\bfseries ARMA}&\multicolumn{1}{c}{\bfseries }&\multicolumn{3}{c}{\bfseries BNDF}&\multicolumn{1}{c}{\bfseries }&\multicolumn{3}{c}{\bfseries DF}&\multicolumn{1}{c}{\bfseries }&\multicolumn{3}{c}{\bfseries SW}\tabularnewline
	 		\cline{3-5} \cline{7-9} \cline{11-13} \cline{15-17}
	 		\multicolumn{1}{c}{ \rule{0pt}{2.5ex}  \vspace{0.02 in}}&\multicolumn{1}{c}{}&\multicolumn{1}{c}{\textbf{N}}&\multicolumn{1}{c}{\textbf{Q1}}&\multicolumn{1}{c}{\textbf{Q2}}&\multicolumn{1}{c}{}&\multicolumn{1}{c}{\textbf{N}}&\multicolumn{1}{c}{\textbf{Q1}}&\multicolumn{1}{c}{\textbf{Q2}}&\multicolumn{1}{c}{}&\multicolumn{1}{c}{\textbf{N}}&\multicolumn{1}{c}{\textbf{Q1}}&\multicolumn{1}{c}{\textbf{Q2}}&\multicolumn{1}{c}{}&\multicolumn{1}{c}{\textbf{N}}&\multicolumn{1}{c}{\textbf{Q1}}&\multicolumn{1}{c}{\textbf{Q2}}\tabularnewline
	 		\hline
	 		\endfirsthead\caption[]{\em (continued)} \tabularnewline
	 		\hline
	 		\multicolumn{1}{c}{\bfseries Model}&\multicolumn{1}{c}{\bfseries }&\multicolumn{3}{c}{\bfseries ARMA}&\multicolumn{1}{c}{\bfseries \rule{0pt}{2.5ex}  \vspace{0.02 in}}&\multicolumn{3}{c}{\bfseries BNDF}&\multicolumn{1}{c}{\bfseries }&\multicolumn{3}{c}{\bfseries DF}&\multicolumn{1}{c}{\bfseries }&\multicolumn{3}{c}{\bfseries SW}\tabularnewline
	 		\cline{3-5} \cline{7-9} \cline{11-13} \cline{15-17}
	 		\multicolumn{1}{c}{ \rule{0pt}{2.5ex}  \vspace{0.02 in} }&\multicolumn{1}{c}{}&\multicolumn{1}{c}{\textbf{N}}&\multicolumn{1}{c}{\textbf{Q1}}&\multicolumn{1}{c}{\textbf{Q2}}&\multicolumn{1}{c}{}&\multicolumn{1}{c}{\textbf{N}}&\multicolumn{1}{c}{\textbf{Q1}}&\multicolumn{1}{c}{\textbf{Q2}}&\multicolumn{1}{c}{}&\multicolumn{1}{c}{\textbf{N}}&\multicolumn{1}{c}{\textbf{Q1}}&\multicolumn{1}{c}{\textbf{Q2}}&\multicolumn{1}{c}{}&\multicolumn{1}{c}{\textbf{N}}&\multicolumn{1}{c}{\textbf{Q1}}&\multicolumn{1}{c}{\textbf{Q2}}\tabularnewline
	 		\hline
	 		\endhead
	 		\hline
	 		\endfoot
	 		AdL&\rule{0pt}{2.5ex}&0.54&0.57&0.75&&0.88&0.84&0.83&&0.62*&0.69&0.95&&0.85&0.72&0.86\tabularnewline
	 		AdP&&0.54&0.60&0.75&&0.89&0.87&0.83&&0.62*&0.72&0.95&&0.85&0.76&0.85\tabularnewline
	 		AdRL&&\textbf{0.53}&0.57&0.74&&\textbf{0.87}&0.83&0.82&&\textbf{0.61*}&0.69&0.94&&\textbf{0.84}&0.72&0.84\tabularnewline
	 		L&&0.54&0.58&0.75&&0.88&0.85&0.83&&0.62*&0.70&0.95&&0.85&0.74&0.85\tabularnewline
	 		L0&&0.76&0.73&0.77&&1.23*&1.07&0.85&&0.86&0.88&0.97&&1.18&0.92&0.87\tabularnewline
	 		L0L1&&0.64&0.61&0.73&&1.04&0.90&0.81&&0.73&0.74&0.93&&1.00&0.77&0.83\tabularnewline
	 		L0L1p&&0.59&0.58&0.75&&0.95&0.85&0.82&&0.67&0.70&0.95&&0.92&0.73&0.85\tabularnewline
	 		L0L2&&0.69&0.61&0.78&&1.13&0.89&0.86&&0.79&0.74&0.98&&1.08&0.77&0.88\tabularnewline
\hline
	 \end{longtable}}
 
	 \subsection{Euro area}
	 \label{appendix:tables:eu}
	 \setlongtables{\scriptsize
	 	\begin{longtable}{lclllclllclllclll}\caption{Spain: PFCE. Relative RMSE of model forecasts during the rolling window pseudo-real-time experiments against four benchmarks: ARMA, BNDF, DF and SW models. The labels N, Q1 and Q2 denotes the forecast horizons -- Nowcast, Forecast-1Q and Forecast-2Q, respectively. The bolded values correspond to the greatest accuracy within the forecast horizon, while the asterix denotes significant performance improvement as suggested by GW test with 5\% significance. \label{es:p3}}  \tabularnewline
	 		\hline\hline
	 		\multicolumn{1}{c}{\bfseries Model}&\multicolumn{1}{c}{\bfseries  \rule{0pt}{2.5ex}  \vspace{0.02 in} }&\multicolumn{3}{c}{\bfseries ARMA}&\multicolumn{1}{c}{\bfseries }&\multicolumn{3}{c}{\bfseries BNDF}&\multicolumn{1}{c}{\bfseries }&\multicolumn{3}{c}{\bfseries DF}&\multicolumn{1}{c}{\bfseries }&\multicolumn{3}{c}{\bfseries SW}\tabularnewline
	 		\cline{3-5} \cline{7-9} \cline{11-13} \cline{15-17}
	 		\multicolumn{1}{c}{}&\multicolumn{1}{c}{ \rule{0pt}{2.5ex}  \vspace{0.02 in} }&\multicolumn{1}{c}{\textbf{N}}&\multicolumn{1}{c}{\textbf{Q1}}&\multicolumn{1}{c}{\textbf{Q2}}&\multicolumn{1}{c}{}&\multicolumn{1}{c}{\textbf{N}}&\multicolumn{1}{c}{\textbf{Q1}}&\multicolumn{1}{c}{\textbf{Q2}}&\multicolumn{1}{c}{}&\multicolumn{1}{c}{\textbf{N}}&\multicolumn{1}{c}{\textbf{Q1}}&\multicolumn{1}{c}{\textbf{Q2}}&\multicolumn{1}{c}{}&\multicolumn{1}{c}{\textbf{N}}&\multicolumn{1}{c}{\textbf{Q1}}&\multicolumn{1}{c}{\textbf{Q2}}\tabularnewline
	 		\hline
	 		\endfirsthead\caption[]{\em (continued)} \tabularnewline
	 		\hline
	 		\multicolumn{1}{c}{\bfseries Model}&\multicolumn{1}{c}{\bfseries  \rule{0pt}{2.5ex}  \vspace{0.02 in} }&\multicolumn{3}{c}{\bfseries ARMA}&\multicolumn{1}{c}{\bfseries }&\multicolumn{3}{c}{\bfseries BNDF}&\multicolumn{1}{c}{\bfseries }&\multicolumn{3}{c}{\bfseries DF}&\multicolumn{1}{c}{\bfseries }&\multicolumn{3}{c}{\bfseries SW}\tabularnewline
	 		\cline{3-5} \cline{7-9} \cline{11-13} \cline{15-17}
	 		\multicolumn{1}{c}{}&\multicolumn{1}{c}{ \rule{0pt}{2.5ex}  \vspace{0.02 in} }&\multicolumn{1}{c}{\textbf{N}}&\multicolumn{1}{c}{\textbf{Q1}}&\multicolumn{1}{c}{\textbf{Q2}}&\multicolumn{1}{c}{}&\multicolumn{1}{c}{\textbf{N}}&\multicolumn{1}{c}{\textbf{Q1}}&\multicolumn{1}{c}{\textbf{Q2}}&\multicolumn{1}{c}{}&\multicolumn{1}{c}{\textbf{N}}&\multicolumn{1}{c}{\textbf{Q1}}&\multicolumn{1}{c}{\textbf{Q2}}&\multicolumn{1}{c}{}&\multicolumn{1}{c}{\textbf{N}}&\multicolumn{1}{c}{\textbf{Q1}}&\multicolumn{1}{c}{\textbf{Q2}}\tabularnewline
	 		\hline
	 		\endhead
	 		\hline
	 		\endfoot
	 		
	 		AdL&\rule{0pt}{2.5ex}&1.07&1.33&1.19&&1.04&1.19&1.14&&1.03&1.10&1.13&&0.97&0.91&0.92\tabularnewline
	 		AdP&&1.03&1.29&1.22&&1.01&1.15&1.17&&1.00&1.06&1.16&&0.94&0.88&0.94\tabularnewline
	 		AdRL&&1.04&1.34&1.19&&1.01&1.20&1.15&&1.00&1.11&1.13&&0.95&0.92&0.92\tabularnewline
	 		L&&1.02&1.26*&1.19&&0.99&1.12&1.14&&0.98&1.04&1.13&&0.93&0.86&0.92\tabularnewline
	 		L0&&1.16&1.52&1.16&&1.13&1.36&1.12&&1.12&1.26&1.10&&1.06&1.04&0.89\tabularnewline
	 		L0L1&&\textbf{0.97}&1.24&1.03&&\textbf{0.95}&1.11&0.99&&\textbf{0.94}&1.03&0.98&&\textbf{0.89}&0.85&0.79*\tabularnewline
	 		L0L1p&&1.03&1.29&1.10&&1.01&1.15&1.06&&1.00&1.07&1.04&&0.94&0.88&0.85\tabularnewline
	 		L0L2&&1.01&\textbf{1.16}&\textbf{0.96}&&0.98&\textbf{1.04}&\textbf{0.92}&&0.97&\textbf{0.96}&\textbf{0.91}&&0.92&\textbf{0.79}&\textbf{0.74*}\tabularnewline
\hline
	 \end{longtable}}
	 
	 \setlongtables{\scriptsize
	 	\begin{longtable}{lclllclllclllclll}\caption{Spain: GFCF. Relative RMSE of model forecasts during the rolling window pseudo-real-time experiments against four benchmarks: ARMA, BNDF, DF and SW models. The labels N, Q1 and Q2 denotes the forecast horizons -- Nowcast, Forecast-1Q and Forecast-2Q, respectively. The bolded values correspond to the greatest accuracy within the forecast horizon, while the asterix denotes significant performance improvement as suggested by GW test with 5\% significance. \label{es:p51g}}  \tabularnewline
	 		\hline\hline
	 		\multicolumn{1}{c}{\bfseries Model}&\multicolumn{1}{c}{\bfseries  \rule{0pt}{2.5ex}  \vspace{0.02 in} }&\multicolumn{3}{c}{\bfseries ARMA}&\multicolumn{1}{c}{\bfseries }&\multicolumn{3}{c}{\bfseries BNDF}&\multicolumn{1}{c}{\bfseries }&\multicolumn{3}{c}{\bfseries DF}&\multicolumn{1}{c}{\bfseries }&\multicolumn{3}{c}{\bfseries SW}\tabularnewline
	 		\cline{3-5} \cline{7-9} \cline{11-13} \cline{15-17}
	 		\multicolumn{1}{c}{}&\multicolumn{1}{c}{ \rule{0pt}{2.5ex}  \vspace{0.02 in} }&\multicolumn{1}{c}{\textbf{N}}&\multicolumn{1}{c}{\textbf{Q1}}&\multicolumn{1}{c}{\textbf{Q2}}&\multicolumn{1}{c}{}&\multicolumn{1}{c}{\textbf{N}}&\multicolumn{1}{c}{\textbf{Q1}}&\multicolumn{1}{c}{\textbf{Q2}}&\multicolumn{1}{c}{}&\multicolumn{1}{c}{\textbf{N}}&\multicolumn{1}{c}{\textbf{Q1}}&\multicolumn{1}{c}{\textbf{Q2}}&\multicolumn{1}{c}{}&\multicolumn{1}{c}{\textbf{N}}&\multicolumn{1}{c}{\textbf{Q1}}&\multicolumn{1}{c}{\textbf{Q2}}\tabularnewline
	 		\hline
	 		\endfirsthead\caption[]{\em (continued)} \tabularnewline
	 		\hline
	 		\multicolumn{1}{c}{\bfseries Model}&\multicolumn{1}{c}{\bfseries  \rule{0pt}{2.5ex}  \vspace{0.02 in} }&\multicolumn{3}{c}{\bfseries ARMA}&\multicolumn{1}{c}{\bfseries }&\multicolumn{3}{c}{\bfseries BNDF}&\multicolumn{1}{c}{\bfseries }&\multicolumn{3}{c}{\bfseries DF}&\multicolumn{1}{c}{\bfseries }&\multicolumn{3}{c}{\bfseries SW}\tabularnewline
	 		\cline{3-5} \cline{7-9} \cline{11-13} \cline{15-17}
	 		\multicolumn{1}{c}{}&\multicolumn{1}{c}{ \rule{0pt}{2.5ex}  \vspace{0.02 in} }&\multicolumn{1}{c}{\textbf{N}}&\multicolumn{1}{c}{\textbf{Q1}}&\multicolumn{1}{c}{\textbf{Q2}}&\multicolumn{1}{c}{}&\multicolumn{1}{c}{\textbf{N}}&\multicolumn{1}{c}{\textbf{Q1}}&\multicolumn{1}{c}{\textbf{Q2}}&\multicolumn{1}{c}{}&\multicolumn{1}{c}{\textbf{N}}&\multicolumn{1}{c}{\textbf{Q1}}&\multicolumn{1}{c}{\textbf{Q2}}&\multicolumn{1}{c}{}&\multicolumn{1}{c}{\textbf{N}}&\multicolumn{1}{c}{\textbf{Q1}}&\multicolumn{1}{c}{\textbf{Q2}}\tabularnewline
	 		\hline
	 		\endhead
	 		\hline
	 		\endfoot
	 		
	 		AdL&\rule{0pt}{2.5ex}&0.97&0.91&0.93&&1.25&0.96&0.95&&1.21&1.13&1.05&&1.02&0.80&0.81\tabularnewline
	 		AdP&&1.02&0.93&0.92&&1.32&0.98&0.94&&1.28&1.15&1.04&&1.07&0.82&0.80\tabularnewline
	 		AdRL&&1.06&0.94&0.91&&1.36&0.99&0.92&&1.32&1.16&1.03&&1.11&0.83&0.79\tabularnewline
	 		L&&0.91&0.89&0.92&&1.17&0.93&0.93&&1.13&1.10&1.04&&0.95&0.78&0.80\tabularnewline
	 		L0&&1.05&0.91&0.94&&1.36*&0.96&0.96&&1.31&1.12&1.07&&1.11&0.80&0.82\tabularnewline
	 		L0L1&&0.95&0.93&0.94&&1.23&0.98&0.95&&1.19&1.15&1.06&&1.00&0.82&0.81\tabularnewline
	 		L0L1p&&0.91&0.91&0.92&&1.17&0.95&0.93&&1.13&1.12&1.03&&0.95&0.79&0.79\tabularnewline
	 		L0L2&&0.96&0.94&0.97&&1.23&0.99&0.98&&1.20&1.16&1.09&&1.01&0.83&0.84\tabularnewline
\hline
	 \end{longtable}}
	 
	 \setlongtables{\scriptsize
	 	\begin{longtable}{lclllclllclllclll}\caption{Spain: Exports. Relative RMSE of model forecasts during the rolling window pseudo-real-time experiments against four benchmarks: ARMA, BNDF, DF and SW models. The labels N, Q1 and Q2 denotes the forecast horizons -- Nowcast, Forecast-1Q and Forecast-2Q, respectively. The bolded values correspond to the greatest accuracy within the forecast horizon, while the asterix denotes significant performance improvement as suggested by GW test with 5\% significance. \label{es:p6}}  \tabularnewline
	 		\hline\hline
	 		\multicolumn{1}{c}{\bfseries Model}&\multicolumn{1}{c}{\bfseries  \rule{0pt}{2.5ex}  \vspace{0.02 in} }&\multicolumn{3}{c}{\bfseries ARMA}&\multicolumn{1}{c}{\bfseries }&\multicolumn{3}{c}{\bfseries BNDF}&\multicolumn{1}{c}{\bfseries }&\multicolumn{3}{c}{\bfseries DF}&\multicolumn{1}{c}{\bfseries }&\multicolumn{3}{c}{\bfseries SW}\tabularnewline
	 		\cline{3-5} \cline{7-9} \cline{11-13} \cline{15-17}
	 		\multicolumn{1}{c}{}&\multicolumn{1}{c}{ \rule{0pt}{2.5ex}  \vspace{0.02 in} }&\multicolumn{1}{c}{\textbf{N}}&\multicolumn{1}{c}{\textbf{Q1}}&\multicolumn{1}{c}{\textbf{Q2}}&\multicolumn{1}{c}{}&\multicolumn{1}{c}{\textbf{N}}&\multicolumn{1}{c}{\textbf{Q1}}&\multicolumn{1}{c}{\textbf{Q2}}&\multicolumn{1}{c}{}&\multicolumn{1}{c}{\textbf{N}}&\multicolumn{1}{c}{\textbf{Q1}}&\multicolumn{1}{c}{\textbf{Q2}}&\multicolumn{1}{c}{}&\multicolumn{1}{c}{\textbf{N}}&\multicolumn{1}{c}{\textbf{Q1}}&\multicolumn{1}{c}{\textbf{Q2}}\tabularnewline
	 		\hline
	 		\endfirsthead\caption[]{\em (continued)} \tabularnewline
	 		\hline
	 		\multicolumn{1}{c}{\bfseries Model}&\multicolumn{1}{c}{\bfseries  \rule{0pt}{2.5ex}  \vspace{0.02 in} }&\multicolumn{3}{c}{\bfseries ARMA}&\multicolumn{1}{c}{\bfseries }&\multicolumn{3}{c}{\bfseries BNDF}&\multicolumn{1}{c}{\bfseries }&\multicolumn{3}{c}{\bfseries DF}&\multicolumn{1}{c}{\bfseries }&\multicolumn{3}{c}{\bfseries SW}\tabularnewline
	 		\cline{3-5} \cline{7-9} \cline{11-13} \cline{15-17}
	 		\multicolumn{1}{c}{}&\multicolumn{1}{c}{ \rule{0pt}{2.5ex}  \vspace{0.02 in} }&\multicolumn{1}{c}{\textbf{N}}&\multicolumn{1}{c}{\textbf{Q1}}&\multicolumn{1}{c}{\textbf{Q2}}&\multicolumn{1}{c}{}&\multicolumn{1}{c}{\textbf{N}}&\multicolumn{1}{c}{\textbf{Q1}}&\multicolumn{1}{c}{\textbf{Q2}}&\multicolumn{1}{c}{}&\multicolumn{1}{c}{\textbf{N}}&\multicolumn{1}{c}{\textbf{Q1}}&\multicolumn{1}{c}{\textbf{Q2}}&\multicolumn{1}{c}{}&\multicolumn{1}{c}{\textbf{N}}&\multicolumn{1}{c}{\textbf{Q1}}&\multicolumn{1}{c}{\textbf{Q2}}\tabularnewline
	 		\hline
	 		\endhead
	 		\hline
	 		\endfoot
	 		
	 		AdL&\rule{0pt}{2.5ex}&0.73&0.78&0.90&&0.84&0.93&0.98&&0.93&0.89&0.91*&&0.79&0.76&0.82\tabularnewline
	 		AdP&&0.72&0.79&0.88&&0.83&0.95&0.97&&0.93&0.91&0.90*&&0.78&0.78&0.81\tabularnewline
	 		AdRL&&0.73&0.78&\textbf{0.88}&&0.84&0.93&\textbf{0.96}&&0.94&0.89&\textbf{0.89*}&&0.79&0.76&\textbf{0.80}\tabularnewline
	 		L&&0.75&0.80&0.89&&0.86&0.96&0.98&&0.96&0.91*&0.91*&&0.81&0.78&0.81\tabularnewline
	 		L0&&0.88&0.91&1.03&&1.01&1.08&1.13&&1.13&1.03&1.05&&0.95&0.89&0.94\tabularnewline
	 		L0L1&&0.85&0.88&0.97&&0.98&1.05&1.07&&1.09&1.01&0.99&&0.92&0.86&0.89\tabularnewline
	 		L0L1p&&0.87&0.86&0.97&&1.00&1.03&1.07&&1.11&0.98&0.99&&0.94&0.84&0.89\tabularnewline
	 		L0L2&&0.92&0.93&1.00&&1.07&1.11&1.10&&1.19&1.06&1.01&&1.00&0.91&0.91\tabularnewline
\hline
	 \end{longtable}}
	 
	 \setlongtables{\scriptsize
	 	\begin{longtable}{lclllclllclllclll}\caption{Spain: Imports. Relative RMSE of model forecasts during the rolling window pseudo-real-time experiments against four benchmarks: ARMA, BNDF, DF and SW models. The labels N, Q1 and Q2 denotes the forecast horizons -- Nowcast, Forecast-1Q and Forecast-2Q, respectively. The bolded values correspond to the greatest accuracy within the forecast horizon, while the asterix denotes significant performance improvement as suggested by GW test with 5\% significance. \label{es:p7}}  \tabularnewline
	 		\hline\hline
	 		\multicolumn{1}{c}{\bfseries Model}&\multicolumn{1}{c}{\bfseries  \rule{0pt}{2.5ex}  \vspace{0.02 in} }&\multicolumn{3}{c}{\bfseries ARMA}&\multicolumn{1}{c}{\bfseries }&\multicolumn{3}{c}{\bfseries BNDF}&\multicolumn{1}{c}{\bfseries }&\multicolumn{3}{c}{\bfseries DF}&\multicolumn{1}{c}{\bfseries }&\multicolumn{3}{c}{\bfseries SW}\tabularnewline
	 		\cline{3-5} \cline{7-9} \cline{11-13} \cline{15-17}
	 		\multicolumn{1}{c}{}&\multicolumn{1}{c}{ \rule{0pt}{2.5ex}  \vspace{0.02 in} }&\multicolumn{1}{c}{\textbf{N}}&\multicolumn{1}{c}{\textbf{Q1}}&\multicolumn{1}{c}{\textbf{Q2}}&\multicolumn{1}{c}{}&\multicolumn{1}{c}{\textbf{N}}&\multicolumn{1}{c}{\textbf{Q1}}&\multicolumn{1}{c}{\textbf{Q2}}&\multicolumn{1}{c}{}&\multicolumn{1}{c}{\textbf{N}}&\multicolumn{1}{c}{\textbf{Q1}}&\multicolumn{1}{c}{\textbf{Q2}}&\multicolumn{1}{c}{}&\multicolumn{1}{c}{\textbf{N}}&\multicolumn{1}{c}{\textbf{Q1}}&\multicolumn{1}{c}{\textbf{Q2}}\tabularnewline
	 		\hline
	 		\endfirsthead\caption[]{\em (continued)} \tabularnewline
	 		\hline
	 		\multicolumn{1}{c}{\bfseries Model}&\multicolumn{1}{c}{\bfseries  \rule{0pt}{2.5ex}  \vspace{0.02 in} }&\multicolumn{3}{c}{\bfseries ARMA}&\multicolumn{1}{c}{\bfseries }&\multicolumn{3}{c}{\bfseries BNDF}&\multicolumn{1}{c}{\bfseries }&\multicolumn{3}{c}{\bfseries DF}&\multicolumn{1}{c}{\bfseries }&\multicolumn{3}{c}{\bfseries SW}\tabularnewline
	 		\cline{3-5} \cline{7-9} \cline{11-13} \cline{15-17}
	 		\multicolumn{1}{c}{}&\multicolumn{1}{c}{ \rule{0pt}{2.5ex}  \vspace{0.02 in} }&\multicolumn{1}{c}{\textbf{N}}&\multicolumn{1}{c}{\textbf{Q1}}&\multicolumn{1}{c}{\textbf{Q2}}&\multicolumn{1}{c}{}&\multicolumn{1}{c}{\textbf{N}}&\multicolumn{1}{c}{\textbf{Q1}}&\multicolumn{1}{c}{\textbf{Q2}}&\multicolumn{1}{c}{}&\multicolumn{1}{c}{\textbf{N}}&\multicolumn{1}{c}{\textbf{Q1}}&\multicolumn{1}{c}{\textbf{Q2}}&\multicolumn{1}{c}{}&\multicolumn{1}{c}{\textbf{N}}&\multicolumn{1}{c}{\textbf{Q1}}&\multicolumn{1}{c}{\textbf{Q2}}\tabularnewline
	 		\hline
	 		\endhead
	 		\hline
	 		\endfoot
	 		
	 		AdL&\rule{0pt}{2.5ex}&0.81&0.79&0.84&&1.22&0.83&0.95&&1.20&0.93&0.91&&1.01&0.73&0.81\tabularnewline
	 		AdP&&\textbf{0.75}&0.81&0.84&&\textbf{1.12}&0.85&0.94&&\textbf{1.10}&0.95&0.91&&\textbf{0.94}&0.74&0.80\tabularnewline
	 		AdRL&&0.80&0.79&0.83&&1.20&0.84&0.94&&1.18&0.93&0.90*&&1.00&0.73&0.80*\tabularnewline
	 		L&&0.80&0.79&0.81&&1.20&0.83&0.92&&1.18&0.92&0.88*&&1.00&0.72&0.78\tabularnewline
	 		L0&&0.80&0.88&0.98&&1.21&0.93&1.10&&1.19&1.04&1.06&&1.01&0.81&0.94\tabularnewline
	 		L0L1&&0.76&0.85&0.88&&1.15&0.90&0.99&&1.13&1.00&0.96&&0.96&0.78&0.85\tabularnewline
	 		L0L1p&&0.80&0.86&0.89&&1.21&0.91&1.00&&1.19&1.01&0.97&&1.01&0.79&0.85\tabularnewline
	 		L0L2&&0.78&0.85&0.86&&1.17&0.90&0.97&&1.15&1.00&0.94&&0.97&0.78&0.83\tabularnewline
\hline
	 \end{longtable}}
	 
	 
	 \setlongtables{\scriptsize
	 	\begin{longtable}{lclllclllclllclll}\caption{Germany: PFCE. Relative RMSE of model forecasts during the rolling window pseudo-real-time experiments against four benchmarks: ARMA, BNDF, DF and SW models. The labels N, Q1 and Q2 denotes the forecast horizons -- Nowcast, Forecast-1Q and Forecast-2Q, respectively. The bolded values correspond to the greatest accuracy within the forecast horizon, while the asterix denotes significant performance improvement as suggested by GW test with 5\% significance. \label{de:p3}} \tabularnewline \tabularnewline
	 		\hline\hline
	 		\multicolumn{1}{c}{\bfseries Model}&\multicolumn{1}{c}{\bfseries  \rule{0pt}{2.5ex}  \vspace{0.02 in} }&\multicolumn{3}{c}{\bfseries ARMA}&\multicolumn{1}{c}{\bfseries }&\multicolumn{3}{c}{\bfseries BNDF}&\multicolumn{1}{c}{\bfseries }&\multicolumn{3}{c}{\bfseries DF}&\multicolumn{1}{c}{\bfseries }&\multicolumn{3}{c}{\bfseries SW}\tabularnewline
	 		\cline{3-5} \cline{7-9} \cline{11-13} \cline{15-17}
	 		\multicolumn{1}{c}{}&\multicolumn{1}{c}{ \rule{0pt}{2.5ex}  \vspace{0.02 in} }&\multicolumn{1}{c}{\textbf{N}}&\multicolumn{1}{c}{\textbf{Q1}}&\multicolumn{1}{c}{\textbf{Q2}}&\multicolumn{1}{c}{}&\multicolumn{1}{c}{\textbf{N}}&\multicolumn{1}{c}{\textbf{Q1}}&\multicolumn{1}{c}{\textbf{Q2}}&\multicolumn{1}{c}{}&\multicolumn{1}{c}{\textbf{N}}&\multicolumn{1}{c}{\textbf{Q1}}&\multicolumn{1}{c}{\textbf{Q2}}&\multicolumn{1}{c}{}&\multicolumn{1}{c}{\textbf{N}}&\multicolumn{1}{c}{\textbf{Q1}}&\multicolumn{1}{c}{\textbf{Q2}}\tabularnewline
	 		\hline
	 		\endfirsthead\caption[]{\em (continued)} \tabularnewline
	 		\hline
	 		\multicolumn{1}{c}{\bfseries Model}&\multicolumn{1}{c}{\bfseries  \rule{0pt}{2.5ex}  \vspace{0.02 in} }&\multicolumn{3}{c}{\bfseries ARMA}&\multicolumn{1}{c}{\bfseries }&\multicolumn{3}{c}{\bfseries BNDF}&\multicolumn{1}{c}{\bfseries }&\multicolumn{3}{c}{\bfseries DF}&\multicolumn{1}{c}{\bfseries }&\multicolumn{3}{c}{\bfseries SW}\tabularnewline
	 		\cline{3-5} \cline{7-9} \cline{11-13} \cline{15-17}
	 		\multicolumn{1}{c}{}&\multicolumn{1}{c}{ \rule{0pt}{2.5ex}  \vspace{0.02 in} }&\multicolumn{1}{c}{\textbf{N}}&\multicolumn{1}{c}{\textbf{Q1}}&\multicolumn{1}{c}{\textbf{Q2}}&\multicolumn{1}{c}{}&\multicolumn{1}{c}{\textbf{N}}&\multicolumn{1}{c}{\textbf{Q1}}&\multicolumn{1}{c}{\textbf{Q2}}&\multicolumn{1}{c}{}&\multicolumn{1}{c}{\textbf{N}}&\multicolumn{1}{c}{\textbf{Q1}}&\multicolumn{1}{c}{\textbf{Q2}}&\multicolumn{1}{c}{}&\multicolumn{1}{c}{\textbf{N}}&\multicolumn{1}{c}{\textbf{Q1}}&\multicolumn{1}{c}{\textbf{Q2}}\tabularnewline
	 		\hline
	 		\endhead
	 		\hline
	 		\endfoot
	 		
	 		AdL&\rule{0pt}{2.5ex}&0.86&1.03&1.04&&0.89&1.09&1.06&&0.86&1.06&1.04&&0.75&0.99&1.01\tabularnewline
	 		AdP&&0.92&1.03&1.05&&0.95&1.09&1.06&&0.91&1.07&1.04&&0.80&1.00&1.02\tabularnewline
	 		AdRL&&0.88&1.05&1.05&&0.91&1.11&1.06&&0.88&1.08&1.04&&0.77&1.01&1.01\tabularnewline
	 		L&&0.91&0.97&0.99&&0.94&1.03&1.00&&0.91&1.00&0.99&&0.79*&0.94&0.96\tabularnewline
	 		L0&&0.97&1.14&1.02&&1.00&1.20&1.03&&0.96&1.17&1.01&&0.84&1.09&0.98\tabularnewline
	 		L0L1&&0.84&0.95&1.08&&0.87&1.00&1.09&&0.84&0.98&1.07&&0.74&0.91&1.04\tabularnewline
	 		L0L1p&&0.94&1.03&1.18&&0.97&1.09&1.19&&0.94&1.07&1.17&&0.82&0.99&1.14\tabularnewline
	 		L0L2&&0.93&0.94&0.96&&0.96&1.00&0.97*&&0.92&0.97&0.96&&0.81&0.91&0.93\tabularnewline
\hline
	 \end{longtable}}
	 
	 \setlongtables{\scriptsize
	 	\begin{longtable}{lclllclllclllclll}\caption{Germany: GFCF. Relative RMSE of model forecasts during the rolling window pseudo-real-time experiments against four benchmarks: ARMA, BNDF, DF and SW models. The labels N, Q1 and Q2 denotes the forecast horizons -- Nowcast, Forecast-1Q and Forecast-2Q, respectively. The bolded values correspond to the greatest accuracy within the forecast horizon, while the asterix denotes significant performance improvement as suggested by GW test with 5\% significance. \label{de:p51g}}\tabularnewline
	 		\hline\hline
	 		\multicolumn{1}{c}{\bfseries Model}&\multicolumn{1}{c}{\bfseries  \rule{0pt}{2.5ex}  \vspace{0.02 in} }&\multicolumn{3}{c}{\bfseries ARMA}&\multicolumn{1}{c}{\bfseries }&\multicolumn{3}{c}{\bfseries BNDF}&\multicolumn{1}{c}{\bfseries }&\multicolumn{3}{c}{\bfseries DF}&\multicolumn{1}{c}{\bfseries }&\multicolumn{3}{c}{\bfseries SW}\tabularnewline
	 		\cline{3-5} \cline{7-9} \cline{11-13} \cline{15-17}
	 		\multicolumn{1}{c}{}&\multicolumn{1}{c}{ \rule{0pt}{2.5ex}  \vspace{0.02 in} }&\multicolumn{1}{c}{\textbf{N}}&\multicolumn{1}{c}{\textbf{Q1}}&\multicolumn{1}{c}{\textbf{Q2}}&\multicolumn{1}{c}{}&\multicolumn{1}{c}{\textbf{N}}&\multicolumn{1}{c}{\textbf{Q1}}&\multicolumn{1}{c}{\textbf{Q2}}&\multicolumn{1}{c}{}&\multicolumn{1}{c}{\textbf{N}}&\multicolumn{1}{c}{\textbf{Q1}}&\multicolumn{1}{c}{\textbf{Q2}}&\multicolumn{1}{c}{}&\multicolumn{1}{c}{\textbf{N}}&\multicolumn{1}{c}{\textbf{Q1}}&\multicolumn{1}{c}{\textbf{Q2}}\tabularnewline
	 		\hline
	 		\endfirsthead\caption[]{\em (continued)} \tabularnewline
	 		\hline
	 		\multicolumn{1}{c}{\bfseries Model}&\multicolumn{1}{c}{\bfseries  \rule{0pt}{2.5ex}  \vspace{0.02 in} }&\multicolumn{3}{c}{\bfseries ARMA}&\multicolumn{1}{c}{\bfseries }&\multicolumn{3}{c}{\bfseries BNDF}&\multicolumn{1}{c}{\bfseries }&\multicolumn{3}{c}{\bfseries DF}&\multicolumn{1}{c}{\bfseries }&\multicolumn{3}{c}{\bfseries SW}\tabularnewline
	 		\cline{3-5} \cline{7-9} \cline{11-13} \cline{15-17}
	 		\multicolumn{1}{c}{}&\multicolumn{1}{c}{ \rule{0pt}{2.5ex}  \vspace{0.02 in} }&\multicolumn{1}{c}{\textbf{N}}&\multicolumn{1}{c}{\textbf{Q1}}&\multicolumn{1}{c}{\textbf{Q2}}&\multicolumn{1}{c}{}&\multicolumn{1}{c}{\textbf{N}}&\multicolumn{1}{c}{\textbf{Q1}}&\multicolumn{1}{c}{\textbf{Q2}}&\multicolumn{1}{c}{}&\multicolumn{1}{c}{\textbf{N}}&\multicolumn{1}{c}{\textbf{Q1}}&\multicolumn{1}{c}{\textbf{Q2}}&\multicolumn{1}{c}{}&\multicolumn{1}{c}{\textbf{N}}&\multicolumn{1}{c}{\textbf{Q1}}&\multicolumn{1}{c}{\textbf{Q2}}\tabularnewline
	 		\hline
	 		\endhead
	 		\hline
	 		\endfoot
	 		
	 		AdL&\rule{0pt}{2.5ex}&0.77&0.95&0.96&&0.95&1.01&1.00&&0.89&1.07&1.02&&0.79&0.93&1.01\tabularnewline
	 		AdP&&0.78&0.94&0.96&&0.97&1.01&1.00&&0.90&1.07&1.02&&0.81&0.93&1.01\tabularnewline
	 		AdRL&&0.82&0.96&0.95&&1.01&1.02&0.98&&0.95&1.08&1.00&&0.84&0.94&0.99\tabularnewline
	 		L&&0.89&0.91&0.95&&1.10&0.97&0.99&&1.03&1.03&1.01&&0.92&0.90&1.00\tabularnewline
	 		L0&&\textbf{0.68}&0.94&0.97&&\textbf{0.85}&1.00&1.00&&\textbf{0.79}&1.06&1.03&&\textbf{0.70}&0.93*&1.02\tabularnewline
	 		L0L1&&0.84&0.93&0.97&&1.04&0.99&1.00&&0.97&1.05&1.03&&0.87&0.92*&1.02\tabularnewline
	 		L0L1p&&0.81&0.94&0.97&&1.00&1.00&1.00*&&0.93&1.06&1.02&&0.83&0.92*&1.01\tabularnewline
	 		L0L2&&0.87&0.91&0.96&&1.08&0.97&0.99&&1.01&1.03&1.02&&0.90&0.90*&1.00\tabularnewline
\hline
	 \end{longtable}}
	 
	 \setlongtables{\scriptsize
	 	\begin{longtable}{lclllclllclllclll}\caption{Germany: Exports. Relative RMSE of model forecasts during the rolling window pseudo-real-time experiments against four benchmarks: ARMA, BNDF, DF and SW models. The labels N, Q1 and Q2 denotes the forecast horizons -- Nowcast, Forecast-1Q and Forecast-2Q, respectively. The bolded values correspond to the greatest accuracy within the forecast horizon, while the asterix denotes significant performance improvement as suggested by GW test with 5\% significance. \label{de:p6}}\tabularnewline
	 		\hline\hline
	 		\multicolumn{1}{c}{\bfseries Model}&\multicolumn{1}{c}{\bfseries  \rule{0pt}{2.5ex}  \vspace{0.02 in} }&\multicolumn{3}{c}{\bfseries ARMA}&\multicolumn{1}{c}{\bfseries }&\multicolumn{3}{c}{\bfseries BNDF}&\multicolumn{1}{c}{\bfseries }&\multicolumn{3}{c}{\bfseries DF}&\multicolumn{1}{c}{\bfseries }&\multicolumn{3}{c}{\bfseries SW}\tabularnewline
	 		\cline{3-5} \cline{7-9} \cline{11-13} \cline{15-17}
	 		\multicolumn{1}{c}{}&\multicolumn{1}{c}{ \rule{0pt}{2.5ex}  \vspace{0.02 in} }&\multicolumn{1}{c}{\textbf{N}}&\multicolumn{1}{c}{\textbf{Q1}}&\multicolumn{1}{c}{\textbf{Q2}}&\multicolumn{1}{c}{}&\multicolumn{1}{c}{\textbf{N}}&\multicolumn{1}{c}{\textbf{Q1}}&\multicolumn{1}{c}{\textbf{Q2}}&\multicolumn{1}{c}{}&\multicolumn{1}{c}{\textbf{N}}&\multicolumn{1}{c}{\textbf{Q1}}&\multicolumn{1}{c}{\textbf{Q2}}&\multicolumn{1}{c}{}&\multicolumn{1}{c}{\textbf{N}}&\multicolumn{1}{c}{\textbf{Q1}}&\multicolumn{1}{c}{\textbf{Q2}}\tabularnewline
	 		\hline
	 		\endfirsthead\caption[]{\em (continued)} \tabularnewline
	 		\hline
	 		\multicolumn{1}{c}{\bfseries Model}&\multicolumn{1}{c}{\bfseries  \rule{0pt}{2.5ex}  \vspace{0.02 in} }&\multicolumn{3}{c}{\bfseries ARMA}&\multicolumn{1}{c}{\bfseries }&\multicolumn{3}{c}{\bfseries BNDF}&\multicolumn{1}{c}{\bfseries }&\multicolumn{3}{c}{\bfseries DF}&\multicolumn{1}{c}{\bfseries }&\multicolumn{3}{c}{\bfseries SW}\tabularnewline
	 		\cline{3-5} \cline{7-9} \cline{11-13} \cline{15-17}
	 		\multicolumn{1}{c}{}&\multicolumn{1}{c}{ \rule{0pt}{2.5ex}  \vspace{0.02 in} }&\multicolumn{1}{c}{\textbf{N}}&\multicolumn{1}{c}{\textbf{Q1}}&\multicolumn{1}{c}{\textbf{Q2}}&\multicolumn{1}{c}{}&\multicolumn{1}{c}{\textbf{N}}&\multicolumn{1}{c}{\textbf{Q1}}&\multicolumn{1}{c}{\textbf{Q2}}&\multicolumn{1}{c}{}&\multicolumn{1}{c}{\textbf{N}}&\multicolumn{1}{c}{\textbf{Q1}}&\multicolumn{1}{c}{\textbf{Q2}}&\multicolumn{1}{c}{}&\multicolumn{1}{c}{\textbf{N}}&\multicolumn{1}{c}{\textbf{Q1}}&\multicolumn{1}{c}{\textbf{Q2}}\tabularnewline
	 		\hline
	 		\endhead
	 		\hline
	 		\endfoot
	 		
	 		AdL&\rule{0pt}{2.5ex}&0.63&0.74&0.97&&1.33&0.97&1.08&&0.91&0.90&1.04&&0.66&0.79&1.07\tabularnewline
	 		AdP&&0.64*&0.75&0.94&&1.35&0.99&1.05&&0.92&0.92&1.01&&0.66*&0.80&1.04\tabularnewline
	 		AdRL&&0.63&0.72&0.97&&1.34&0.96&1.08&&0.92&0.89&1.04&&0.66&0.78&1.07\tabularnewline
	 		L&&0.65&0.82&0.98&&1.38&1.08&1.09&&0.95&1.00&1.05&&0.68&0.88&1.08\tabularnewline
	 		L0&&0.80&0.75&0.95&&1.69*&0.99&1.06&&1.16&0.92&1.02&&0.83&0.80&1.05\tabularnewline
	 		L0L1&&0.69&\textbf{0.72}&0.89&&1.46&\textbf{0.95}&0.99&&1.00&\textbf{0.88}&0.95&&0.72&\textbf{0.77}&0.98\tabularnewline
	 		L0L1p&&0.69&0.74&\textbf{0.88}&&1.46&0.97&\textbf{0.98}&&1.00&0.90&\textbf{0.95}&&0.72&0.79&\textbf{0.97}\tabularnewline
	 		L0L2&&0.72&0.79&0.91&&1.52&1.05&1.02&&1.04&0.97&0.98&&0.75&0.85&1.01\tabularnewline
\hline
	 \end{longtable}%
}

\end{appendices}

\end{document}